\documentclass{emulateapj}

\def\etal{{\it et al.\ }} 
\def\eg{{\it e.g.\ }}

\def\p3m{P${}^3$M} 
\def\ap3m{AP${}^3$M} \def\-{{\em{---}}}

\newcommand{\be}{\begin{equation}}  \newcommand{\ba}{\begin{eqnarray}}
\newcommand{\ee}{\end{equation}}  \newcommand{\ea}{\end{eqnarray}}

\newcommand{\Metal}{_{\mathrm{metal}}}
\newcommand{\Bubble}{_{\mathrm{bbl}}}
\newcommand{\Inj}{_{\mathrm{inj}}}
\newcommand{\Evac}{_{\mathrm{evac}}}

 \newcommand{\bi}{\begin{itemize}}
\newcommand{\ei}{\end{itemize}} 
 
\newcommand{\Myr}{\,\textrm{Myr}}

\def\lesssim{\mathrel{\hbox{\rlap{\hbox{\lower4pt\hbox{$\sim$}}}\hbox{$<$}}}}
\def\gtrsim{\mathrel{\hbox{\rlap{\hbox{\lower4pt\hbox{$\sim$}}}\hbox{$>$}}}}
\def\apj{{ApJ}}

\begin{document}


\lefthead{Scannapieco \& Br\" uggen}  \righthead{}

\title{Subgrid Modeling of AGN-Driven Turbulence in Galaxy Clusters}

\author{Evan Scannapieco\altaffilmark{1} \& Marcus Br\"
uggen\altaffilmark{2}} \altaffiltext{1}{School of Earth and Space
Exploration,  Arizona State University, P.O.  Box 871404, Tempe, AZ,
85287-1404.}  \altaffiltext{2}{Jacobs University Bremen, P.O. Box
750\,561, 28725 Bremen, Germany}

\begin{abstract} 

Hot, underdense bubbles powered by active galactic nuclei (AGN) are
likely to play a key role in halting catastrophic cooling in the
centers of cool-core galaxy clusters.  We present three-dimensional
simulations that capture the evolution of such bubbles, using an
adaptive-mesh hydrodynamic code, FLASH3, to which we have added a
subgrid model of turbulence and mixing.  While pure-hydro simulations
indicate that AGN bubbles are disrupted into resolution-dependent
pockets of  underdense gas, proper modeling of subgrid turbulence
indicates that this a poor approximation to a turbulent cascade that
continues far beyond the resolution limit.  Instead, Rayleigh-Taylor
instabilities act to effectively mix the heated region with its
surroundings, while at the same time preserving it as a coherent
structure, consistent with observations.  Thus bubbles are transformed
into hot clouds  of mixed material as they move outwards in the
hydrostatic intracluster medium (ICM), much as large airbursts lead to
a distinctive  ``mushroom cloud'' structure as they rise in the
hydrostatic atmosphere of Earth.  Properly capturing the evolution of
such clouds has important implications for many ICM properties.  In
particular, it significantly  changes the impact of AGN
on the distribution of entropy and metals in cool-core clusters such
as Perseus.

\end{abstract}


\section{Introduction}

The X-ray and abundance  profiles  of the hot, diffuse
medium in galaxy clusters are observed to  be bimodal (Fukazawa
\etal 2000; Matsushita \etal 2002;  Schmidt \etal 2002; Churazov \etal
2003; De Grandi \etal 2004). Strong intracluster medium (ICM)
abundance gradients are
associated with cool-core clusters with a peak in their central X-ray
surface brightness distributions, and nearly uniform abundance
profiles are associated with non cool-core clusters.  Furthermore,
these metallicity distributions are much broader than the associated
galaxy light, indicating that significant mixing has occurred (\eg
Churazov \etal  2003; Chandran 2005; David \& Nulsen 2008).

At the same time, the nature of cool-core clusters remains uncertain.
Although strong X-ray emission indicates that the central gas is
cooling rapidly, the deficit of star formation and  $<$ 1 keV gas (\eg
Fabian 1994; Tamura \etal 2001) means that radiative cooling must be
balanced by an unknown energy source.  Currently, the most successful
model for achieving this balance relies on heating from a central AGN,  yet the
details of this process are poorly understood (\eg Loken \etal 1995;
Br\" uggen \& Kaiser 2002; Reynolds \etal 2002; Brighenti \& Mathews
2006; Brunetti \& Lazarian 2007).

While AGN are observed to drive large bubbles filled with relativistic
particles (\eg Boehringer \etal 1993; Carilli \etal 1994; McNamara
\etal 2000; Blanton \etal 2001; Finoguenov \& Jones 2001; Nulsen \etal
2005),  the  synchrotron radiation emitted by the electrons in these
regions fades after about $10^8$ years, becoming extremely difficult
to detect. Moreover, the  corresponding depressions in the X-ray
surface brightness are only visible near the center of the cluster,
where the contrast is large. Thus, it is unclear how far these
structures rise in the cluster.  Furthermore, AGN have been observed
to induce shocks and/or sonic motions in the ICM
that are believed to eventually dissipate their energy into this gas
(Fabian \etal 2003; Kraft \etal 2003; Ruszkowski \etal 2004; McNamara
\etal 2005), although the impact of the resulting heating is difficult
to quantify observationally.

The presence of AGN-heated cavities in galaxy clusters has raised a number
of questions.  These buoyant bubbles are
unstable  to the Rayleigh-Taylor (RT) instability,  which occurs
whenever a fluid  is accelerated or supported against gravity by a
fluid of lower density.  In any ideal hydrodynamic model, the cavities
must be inflated supersonically or else they would be destroyed by RT
instabilities faster than they are inflated (Reynolds \etal 2005).
Curiously, the strong and hot ICM shocks that are expected around the
active cavities are absent, and, instead, many cavities are surrounded
by shells of gas that is cooler than the ambient ICM (Nulsen \etal 2002).
 Secondly, these
cavities appear to be intact even after inferred ages of several
$10^8$ yrs, as for example the outer cavities in Perseus (Nulsen \etal
2002). However, hydrodynamic simulations fail to reproduce the
observed morphology as the RT and other instabilities shred the bubbles in
a relatively short time (Robinson \etal 2004; Reynolds \etal 2005,
Br\"uggen \etal 2005a), although this time can  be extended somewhat
by a more detailed
treatment of bubble inflation (Pizzolato \& Soker 2006; Sternberg \etal 2008;
Sternberg \& Soker 2008a,b).

A consequence of the evolution of such cavities is the
production of turbulence, which is likely to be pervasive in the ICM 
and play a crucial role in the structure of cool-core clusters (\eg Schuecker 
\etal 2004).  Turbulence occurs in any case in which the Reynolds
number, $Re$, is greater than $\approx 1000$,  
where  $Re \equiv d \, v /\nu,$  $d$ is
the characteristic scale of the flow instability, $v$ is its
characteristic velocity, and $\nu$ is the kinetic viscosity  of the
fluid.   While there have been some suggestions that the ICM may have
a non-negligible viscosity (Ruszkowski \etal 2004;
Reynolds \etal 2005), this quantity remains unknown  because the
physics of such dilute and magnetized plasmas is poorly constrained.
In particular,  even minute magnetic fields lead to small proton
gyroradii that suppress viscosity efficiently.
However, it has been pointed out that the exponential divergence of
neighboring field lines in a tangled magnetic field may lead to only 
modest suppression (\eg Narayan \& Medvedev 2001). On the other
hand, Rebusco \etal (2005) showed that the turbulent  scales and
velocities required to spread metals in the Perseus cluster closely
correspond to those necessary to balance cooling, and similar results
have been recently obtained for several other clusters (Rebusco \etal
2006).  This turbulence has important implications for  particle
acceleration and the mixture and transport of gas energy and  heavy 
elements.

Observationally, there are several circumstantial clues about the
nature of turbulence in the ICM, such as the pressure distribution in
the Coma cluster (Schuecker \etal 2004), the lack of resonant
scattering in the 6.7 iron K$\alpha$ line in the Perseus cluster
(Churazov \etal 2004) and the Faraday rotation map of the Hydra
cluster (Vogt \& Ensslin 2005, see also Iapichino \& Niemeyer 2008).
Turbulence has also been invoked to explain the non-thermal emission
in clusters (\eg Brunetti \& Lazarian
2007), and Doppler shifts from such bulk motions are likely
to be directly detectable with the next generation of X-ray observatories,
such as {\em Constellation-X}, 
with an envisaged spectral resolution of 1-2 eV (\eg Br\"uggen \etal 2005b).

Finally, our understanding of AGN-driven turbulence in cool-core
clusters is complicated both by a variety of other possible sources of
turbulence and competing physical effects.  Clusters form through
the accretion of smaller structures  and this infall 
can generate turbulence (see \eg Takizawa
2005). Episodes of active merging are also expected to produce
turbulence, as seen in simulations, (\eg  Norman \& Bryan 1999; Ricker
\& Sarazin 2001). Within clusters, the motion of galaxies
can also produce wakes that are likely to generate turbulent motions
(Bregman \& David 1989; Kim 2007), and  microphysical processes,
often described in terms of conduction and viscosity, may also play
important roles (\eg Narayan \& Medvedev 2001; Voigt \& Fabian 2004;
Ruszkowski \etal 2004; Sternberg \etal 2007).  In summary, cool-core 
clusters are a mystery
that is carefully observed, poorly understood, and closely tied up
with AGN-driven turbulence.

It is with this in mind that we have carried out detailed simulations
of AGN-driven turbulence in a cool-core cluster using the adaptive
mesh refinement code, FLASH3.  While the direct simulation of
turbulence is extremely challenging, computationally expensive, and
dependent on resolution  (\eg  Glimm \etal 2001), its
behavior can be approximated to a good degree of accuracy by adopting
a sub-grid approach. In this case the flow is decomposed into mean and
fluctuating components, which provides a systematic framework for
deriving a set of turbulence equations.

There are several types of models that are able to describe 
such fluctuations arising from  the RT instability as
well as the Richtmyer-Meshkov instability, which occurs when a shock
hits a medium of varying acoustic impedance. The simplest such 
model consists of ordinary differential equations for the mixing
region (\eg Alon 1995; Chen \etal 1996; Ramshaw 1998), describing for
example, the amplitude of the bubble by balancing inertia, buoyancy,
and drag forces.  While these yield the right growth rates,
they fail when there are multiple interfaces  and are not readily
extended to two and more dimensions. Although these problems can be
addressed with multifluid models (Youngs 1989), such models are
complicated, numerically expensive, and sometimes unstable.

A second class of models evolves the turbulent kinetic energy per unit
mass and its dissipation rate. Such ``two-equation turbulence
models'' developed for unstable shear flows postulate a turbulent
viscosity, a Reynolds stress, and a dissipation term (\eg Llor
2003). However, the usual Reynolds stress terms must be modified in
the presence of shocks, and modeling the RT and RM instabilities requires a
buoyancy term that depends on the amplitude and the wavelength.

Recently, DiMonte \& Tipton (2006, hereafter DT06), described a 
sub-grid model that is especially suited to capturing the
buoyancy-driven turbulent evolution of AGN bubbles. The model
captures the self-similar growth of the RT and RM instabilities by
augmenting the mean hydrodynamics equations with evolution equations
for the turbulent kinetic energy per unit mass and the
scale length of the dominant eddies. The equations are based on
buoyancy-drag models for RT and RM flows, but
constructed with local parameters so that they can be applied to
multidimensional flows with multiple materials.  The model
is self-similar, conserves energy, preserves Galilean
invariance, and works in the presence of shocks, and although it contains
several unknown coefficients, these are determined by comparisons with
analytic solutions, numerical simulations, and experiments.

Here we implement the DT06 model into FLASH3 to: 1.) examine the
impact of turbulence on the morphology and stability of AGN-driven
bubbles as observed in nearby clusters (\eg Fabian 2006); 2.)
quantify turbulence, comparing it to present indirect entropy and
metal distribution constraints (\eg Inogamov \& Sunyaev 2003) and
making predictions for radial profiles,  as directly measurable  by
future linewidth studies.  Our goal is to focus on better
understanding the basic case of purely AGN-driven turbulence in an
inviscid, unmagnetized ICMm and to this end, we follow the
model of Roediger \etal (2007; hereafter R07), in which the ICM is
described by a spherically-symmetric profile fit to X-ray
observations of the Perseus cluster, and feedback is modeled by
periodically injecting energy into the center of this distribution.

The structure of this work is as follows.  In \S2 we give an overview
of the FLASH3 code and our implementation the DT06 subgrid-turbulence
model within it.  In \S3 we present tests of our implementation
against analytic solutions.  In \S4 we discuss our
modeling of the galaxy cluster and energy input by the central AGN. In
\S5 we present the results from our simulations and discuss
their observational consequences.  Our conclusions are summarized in
\S 6.

\vspace{2cm}

\section{Numerical Modeling}

\subsection{FLASH}

All simulations in this study were performed with the multidimensional
adaptive-mesh refinement (AMR) code FLASH3 (Fryxell \etal 2000).  FLASH3 is
a modular block-structured code, which is parallelized using the
Message Passing Interface (MPI) library. It advances the equations of
inviscid hydrodynamics by solving the Riemann problem on a Cartesian
grid using a directionally-split Piecewise-Parabolic Method (PPM) solver
(Colella \& Woodward 1984; Colella \& Glaz 1984; Fryxell, M\" uller, \& Arnett 1989),
a higher-order version of the method developed in Godunov (1959). 
This approach uses a
monotonicity constraint rather than an artificial viscosity to control
oscillations near discontinuities. However, the discretisation of
the equations still leads to a numerical viscosity, as discussed
in further detail below.  Finally, FLASH3 offers the
opportunity to advect mass scalars along with the gas density, a
property that we utilize both in
advancing the variables in our subgrid-turbulence model
and capturing the evolution of metals.

\subsection{Turbulence Model and Numerical Implementation}

To capture the development of AGN-driven turbulence and its impact on
the intracluster medium, we make use of the subgrid model
developed in DT06, in which the Navier-Stokes
equations are extended with a turbulent viscosity,  $\mu_t$, that
depends on a turbulent eddy size, $L,$ and a turbulent kinetic energy
per unit mass, $K$.  The flow is divided up into a mean component and
fluctuating component, such that the total velocity ${\bf u}$ is given
by 
\be 
{\bf u} = {\bf \tilde u} + {\bf u'}, 
\ee 
where ${\bf \tilde
u}$ is the mass averaged mean velocity ${\bf \tilde  u} \equiv
{\overline{\rho {\bf u}}}/{\overline \rho},$  and ${\overline{\rho
{\bf u'}}}=0,$ where $\rho$ is the mass density and the overbar 
indicates an ensemble average over many realization of the flow.
To leading order,
this expansion gives the following evolutionary equations: \ba
\frac{\partial {\bar \rho}}{\partial t} + \frac{\partial {\bar \rho}
\tilde u_j}{\partial x_j} &=& 0,
\label{eq:rho}\\
\frac{\partial {\bar \rho} \tilde u_i}{\partial t} + \frac{\partial
{\bar \rho} \tilde u_i \tilde u_j}{\partial x_j} &=& -\frac{\partial
P}{\partial x_i} -\frac{\partial R_{i,j}}{\partial x_j},
\label{eq:u}\\
\frac{\partial {\bar \rho} E}{\partial t} + \frac{\partial {\bar \rho}
E \tilde u_j}{\partial x_j} &=&  \frac{\partial}{\partial x_j} \left(
\frac{\mu_t}{N_E}  \frac{\partial E}{\partial x_j}\right)
-\frac{\partial P \tilde u_j}{\partial x_j} - S_K, \nonumber
\label{eq:E} 
\ea where $t$ and ${\bf x}$ are time and position variables, $\bar
\rho({\bf x},t)$ is the average density field,  $\tilde u_i({\bf
x},t)$  is the mass-averaged mean-flow velocity field in the $i$
direction, $P({\bf x},t)$ is the mean pressure, and $E({\bf x},t)$ is
the mean
internal energy per unit mass. Subgrid turbulence affects these
mean-flow quantities through the explicit source term, $S_K$, the
Reynolds stress tensor $R_{i,j},$ and the turbulent viscosity, which
is scaled in the energy equation by $N_E$. In the case in which
multiple fluids are considered, these equations are supplemented by a
mass-fraction equation: \be \frac{\partial {\bar \rho F_r}}{\partial
t} + \frac{\partial {\bar \rho F_r} \tilde u_j}{\partial x_j}=
\frac{\partial}{\partial x_j} \left(  \frac{\mu_t}{N_F}
\frac{\partial F_r}{\partial x_j}\right)
\label{eq:massfrac},\\
\ee where $F_r$ is the mass fraction of species $r$ in a given zone,
and $N_F$ is a scale factor.

The turbulence quantities that appear in these equations are
calculated from evolution equations for $L$ and $K$.  The eddy scale
$L$ must be evolved because the buoyancy-driven RT and RM
instabilities depend on the eddy size, which is expected to grow
self-similarly. Simple equations that include
diffusion, production, and compression are given by
\be 
\frac{\partial \bar \rho L}{\partial t} + \frac{\partial \bar
\rho L \tilde u_j}{\partial x_j} = \frac{\partial}{\partial x_j}
\left(  \frac{\mu_t}{N_L}  \frac{\partial L}{\partial x_j}\right) +
\bar \rho V + C_C \bar \rho L \frac{\partial \tilde u_i}{\partial x_i}
\label{eq:L},
\ee 
and 
\be 
\frac{\partial \bar \rho K}{\partial t} + \frac{\partial
\bar \rho K \tilde u_j}{\partial x_j} = \frac{\partial}{\partial x_j}
\left(  \frac{\mu_t}{N_K}  \frac{\partial K}{\partial x_j}\right) -
R_{i,j} \frac{\partial \tilde u_i}{\partial x_j} + S_K,
\label{eq:K}
\ee 
where $N_K$, $N_L$, $C_C$ are constants.

In the $L$ equation the three terms on the right hand 
represent, respectively: turbulent diffusion, growth of
eddies through turbulent motions, and growth of eddies due to motions
in the mean flow.  In the $K$ equation the three terms on the right
hand side represent, respectively: turbulent diffusion, the work
associated with the turbulent stress, and a source term $S_K$ that
contains both Rayleigh-Taylor and Richtmyer-Meshkov contributions,
which also appears in the internal energy equation to conserve energy.

The key source term for the RT and RM instabilities is based on the 
successful buoyancy-drag model, namely
\be 
S_K = \bar \rho V \left[ C_B A_i g_i - C_D \frac{V^2}{L} \right],
\label{eq:sk}
\ee 
where the coefficients $C_B$ and $C_D$ are fit to
experiments. Physically, $C_B$ describes turbulent
entrainment, which reduces the density contrast, and the drag
coefficient $C_D$ describes the dissipation of turbulent energy when
the average scale is characterized by $L$.
Moreover, $V \equiv \sqrt{2 K}$ is the
average turbulent velocity, $g_i \equiv - (1/\rho) \partial P/\partial
x_i$ is the gravitational acceleration, and $A_i$ is the Atwood number in the
$i$ direction. In the code, we  determine this as
\be
 A_i = \frac{\bar \rho_+ - \bar \rho_-}{\bar \rho_+ +
\bar \rho_-} + C_A \frac{L}{\bar \rho + L |\partial \bar \rho/\partial
x_i|}  \frac{ \partial \bar \rho}{\partial x_i},
\label{eq:Ai}
\ee 
where $C_A $ is a constant and $\bar \rho_-$ and $\bar \rho_+$ are the densities on
the back and front boundaries of the cell in the $i$ direction.

As the inclusion of shear-driven turbulence in the DM06 model is still
in the process of development, for the Reynolds stress tensor we
consider only the turbulent pressure
\be R_{i,j} = C_P \delta_{i,j} \bar \rho K,
\label{eq:rij}
\ee 
where $\delta_{i,j}$ is the Kroniker delta
and $C_P$ is a constant.  Thus, although we expect buoyancy-driven
turbulence to dominate ICM mixing by hot bubbles, our results 
nevertheless represent a lower-limit that does not
include shear-driven turbulence such as that arising from 
the Kelvin-Helmholtz instability (Helmholtz 1868; Kelvin 1871).

Finally, the turbulent viscosity is calculated as  \be \mu_t = C_\mu
\bar \rho L V,
\label{eq:mut}
\ee where $C_\mu$ is a constant.  A list of the model coefficients and
their values is given in Table \ref{table:DT}, in which we also summarize
the effect of each constant on the model.

\begin{table}
\caption{Summary of coefficients in the Turbulence model.   Constants
that are fit to experiment appear with error bars, and in those cases
we take the central value for this study.}
\label{tab:ICM}
\centering
\begin{tabular}{lll} 
\hline 
Parameter & Value & Effect \\
\hline 
$N_L$ & $0.5 \pm 0.1$ & Diffusion of $L$ \\ 
$N_K$ & $1.0 \pm 0.2$ &
Diffusion of $K$ \\ $N_F$ & $1.0 \pm 0.2$ & Diffusion of Species $F$\\
$N_E$ & $1.0 \pm 0.2$ & Diffusion of $E$ \\ $C_A$ & $2.0$ & Turbulent
Atwood Number\\ $C_B$ & $0.84 \pm 0.11$ & Buoyancy-Driven Turbulence
\\ $C_C$ & $1/3$           &  Compression of $L$\\ $C_D$ & $1.25 \pm
0.4$  & Drag term for $K$\\ $C_P$ & $2/3$           & Turbulent
Pressure \\ $C_\mu$ & $1.0$         & Turbulent Viscosity \\ \hline
\label{table:DT}
\end{tabular}
\end{table}

Our numerical implementation of these equations is divided into three
steps that are carried out after the main hydro step in FLASH3, which advects
all the variables above, including $K$ and $L$.   In the first step, we
implement the $\partial \tilde u_i/ \partial x_i$ terms in eqs.\
(\ref{eq:u}), (\ref{eq:L}), and (\ref{eq:K}) explicitly.    In the
second step, we: (i) compute $V$ as $\sqrt{2 K}$, (ii) use a leapfrog
technique to add the source term to $V$ as $S_K/\rho V$ along with the
$\bar \rho V$ to the $L$ equation, and then (iii) write $V$ back to
the $K$ array as $K = V^2/2.$  Finally, in the third step, we
calculate the turbulent viscosity and use this to implement the
diffusive mixing terms in eqs.\ (\ref{eq:E}), (\ref{eq:L}), and
(\ref{eq:K}) explicitly.   This final step requires us to impose an
additional constraint on the minimum times step of  $dt \leq
(\Delta^2/\mu_t)/4$ where $(\Delta)$ is the minimum of  $dx$, $dy$,
and $dz$ in any given zone.  This diffusive constraint must be
satisfied for all zones in the simulation.

\section{Tests of Our Subgrid-Turbulence Model}

\subsection{Rayleigh-Taylor Shock-Tube Test}

\begin{figure}
\centerline{\includegraphics[height=8.5cm]{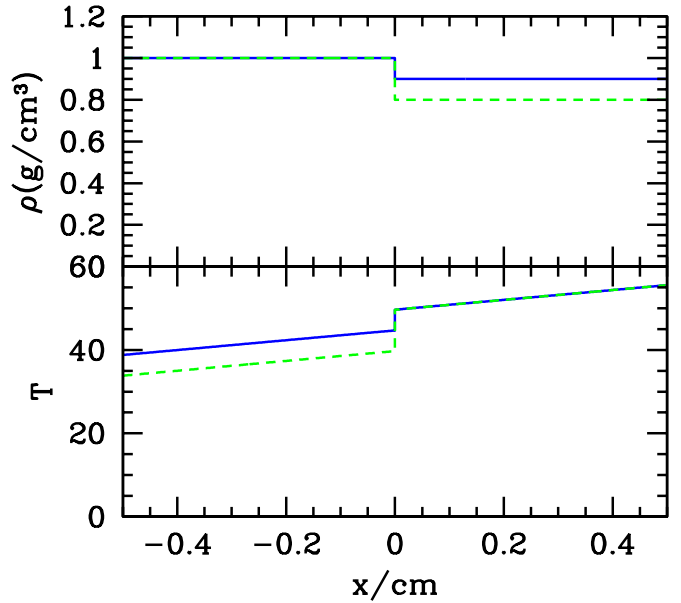}}
\caption{Initial set-up for our Rayleigh-Taylor test simulations.
{\em Top:} Initial density profile  in the $\rho_2 = 0.9$ case (solid
line), and in the $\rho_2 = 0.8$ case (dashed line).   {\em Bottom:}
Temperature profiles in both cases, with lines as in the upper panel.}
\label{fig:profiles}
\end{figure}

As a test of our implementation, we recreated the Rayleigh-Taylor test
problem described in section V of DT06.  We considered a $1$ cm
region that was filled with two $\gamma = 5/3$ ideal fluids with
constant densities: $\rho_1 =  1$ g/cm$^3$ from $x= -0.5$ to $x=0,$
and $\rho_2 = 0.9$ from $x=0$ to $x=0.5$.  The region was placed in a
gravitational field that pointed in the $x$ direction with an acceleration of
$9.8 \times 10^8$ cm/s$^2$, or $10^6$ times the Earth's gravity, and
the temperature profile was chosen such that the overall distribution
was in hydrostatic balance and at $x=0$ the temperature of the lower
density fluid  was $T_2 = 50$ K. Although this is a one-dimensional
problem, as a check of our implementation our test simulations were
carried out in 2 dimensions in a 50 ``block'' by 1 ``block'' region,
where a FLASH3 block represents  $8 \times 8$ simulation
cells.  Finally, the code was allowed to place up to 3 refinements,
such that $l_{\rm refine}$ was 4, based on the density and pressure
profile, which lead to an initial $dx$ near the interface of $1\, {\rm
cm}/50/8/2^4=  3.1 \times 10^{-4}$ cm, and a minimum resolution of
$2.5 \times 10^{-3}$ cm.  This initial profile is
shown in Figure \ref{fig:profiles}.

\begin{figure}
\centerline{\includegraphics[height=11.0cm]{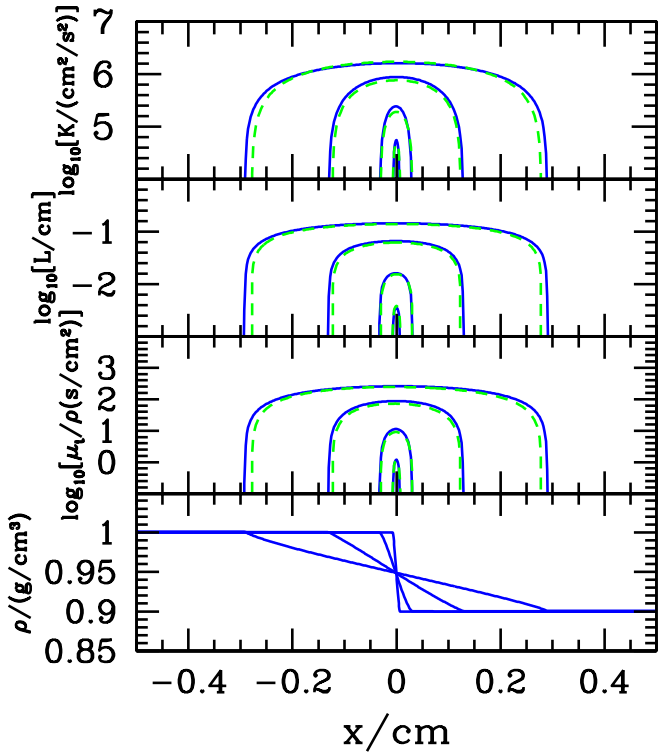}}
\caption{Evolution in the $\rho_2 = 0.9$ case.   {\em Top:} Profiles
of the turbulent kinetic energy per unit mass at 50 $\mu$s,
100 $\mu$s, 200 $\mu$s, and 300 $\mu$s. In each case the solid line
gives the simulation results and the dashed line gives the analytic
solution.   {\em Second panel:} Profiles of the turbulent length
scales from the simulation (solid) and the analytic solution (dashed),
with times as in the upper panel.  {\em Third Panel:} Profiles of the
simulated  turbulent viscosity per unit mass (solid) as compared to the
analytic values (dashed). {\em Bottom:} Simulated density profiles with
times as in the other panels.\\}
\label{fig:xtest0.9}
\end{figure}

As described in DT06, the growth of turbulence at the interface in
this Rayleigh-Taylor unstable interface has the following analytic
solution: \ba L(x,t) &=& L(t,0) \left[1 - x^2/h(t)^2 \right]^{1/2},
\label{eq:Lana}\\
K(x,t) &=& K(t,0) \left[1 - x^2/h(t)^2 \right],
\label{eq:Kana}
\ea where $h(t) = \alpha_b A(0) t^2$  is a scale length for the
interpenetrating fluid (with $\alpha_b \equiv C_A C_B/[8(1+2.C_D)] =
0.06$), and $L(t,0) = h(t)/2$ and $K(t,0) = (dh/dt)^2/2$ are the
turbulent length scale and turbulent kinetic energy per unit mass at
the interface.   Following DT06, we initialize our simulation at a
time of $10 \mu$s and set $L(0,10 \mu {\rm s}) = h(10 \mu {\rm
s})^2/dx$ in the  zones at either side of the interface. The resulting
$K$, $L$, and density profiles are shown in Figure \ref{fig:xtest0.9}.

At all times, and for all quantities of interest, our implementation
reproduces the analytic solution with a high degree of accuracy.   As
expected, the kinetic energy per unit mass, turbulent viscosity, and
the width of the turbulent layer (taken to be the scale at which $K$
reaches a 1/10 of its maximum value) all increase as $\propto t^2$,
which eventually leads to rapid mixing between the two materials.
Furthermore, although our explicit implementation of turbulent
diffusion requires us to  impose a time step $\propto dx^2/\mu_t$,
this works well in concert with the AMR hydro solver.  As $\mu_t$
increases near the interface, density gradients are smoothed, allowing
the code to derefine these boundaries.  Thus the diffusive time step
remains greater than or comparable to the one required by the Courant
condition for most of the simulation, dropping only to a minimum value
of about $1/6$ of the Courant time step at the very latest  times,
when the entire simulation volume has moved to the lowest allowed
refinement  value.

\begin{figure}
\centerline{\includegraphics[height=11.0cm]{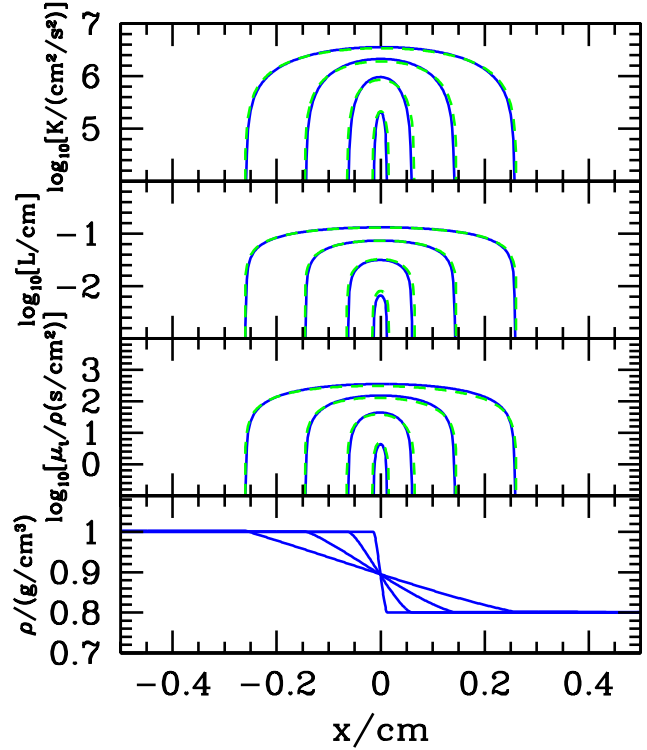}}
\caption{Evolution in the $\rho_2 = 0.8$ case.   {\em Top:} Profiles
of the turbulent kinetic energy per unit mass at times of 50 $\mu$s,
100 $\mu$s, 150 $\mu$s, 200 $\mu$s, where again the solid lines give
the simulation results and the dashed lines give the analytic
solution.   {\em Second Panel:} Profiles of the turbulent length
scales with lines and times as in the upper panel.  {\em Third
Panel:} Profiles of the simulated and analytic turbulent viscosity
per unit mass.
{\em Bottom:} Simulated density profiles.}
\label{fig:xtest0.8}
\end{figure}

In Figure \ref{fig:xtest0.8} we show the results of a similar test in
which the density on the right hand side of the volume decreased to
$0.8$ g/cm$^3$, doubling the growth rate of the turbulent layer.
Again, we find good agreement between our numeric approach and the
analytic results, and again derefinements at late times keep the the
diffusive $dt$ to manageably large values.  We repeated this
2D simulation with the $x$ and $y$ coordinates interchanged, and
carried out a 3D simulation in which gravity was pointed in the $z$
direction.  In both cases we achieved results identical to the ones
presented above.

Finally, we carried out comparison runs in which we initialized the
simulation as in the $x$-direction test runs, increased the resolution
in the $y$ direction to 4 or 8 blocks, and turned off the subgrid
turbulence model.  In this case, the fluid remained  stationary until
perturbations from numerical errors grew to the point that fingers
formed between the two materials.  The delay in the onset of this stage
was dependent on the resolution  in the $y$ direction, and after this
onset, the region containing interpenetrating fingers grew roughly as
$t^2$.

\subsection{Richtmyer-Meshkov Test}

Next, we tested the ability of our code to simulate  Richtmyer-Meshkov
(RM) amplified turbulence (Richtmyer 1960; Meshkov 1969),  which
occurs when a shock encounters a discontinuity in acoustic impedance
such as a density step.  In this case, the shock accelerates  the
interface to a velocity $v_{\rm int},$  which amplifies the initial
perturbations by sending the low-density regions running out ahead of
the higher-density regions (\eg Mikaelian 1989; Alon \etal 1995;
Dimonte 1999; Holmes \etal 1999).  As the perturbations grow and
become nonlinear, the growth  rate decays away in time, leading to a
bubble  amplitude that grows as $h_b \propto (v_{\rm int}
t)^\theta_b,$ where $\theta_b \approx 0.25 \pm 0.05$.

To study the ability of our implementation to capture this
instability, we reproduced the
shock-tube test problem described in section VI of DT06.  In this
case, we considered a $15$ cm long region, spanning from $x=-7$ to
$x=8$.  The rightmost section of the simulation volume, from $x=0$ to
$x=8$, was filled with a stationary gas with $\rho=0.667$ g/cm$^3$ and
$T=2.16$ K.  The center region of the simulation, from $x=-6.9$ to
$x=0$ was filled with a second stationary gas with $\rho=1$ and
$T=1.44$ K.  Both gases has a polytropic index of $\gamma =5/3$. Finally,
the region from $x=-7$ to $x=-6.9$, was filled with a $\rho = 1.81$
g/cm$^3$, $T=2.27$ K, $\gamma =5/3$ gas flowing in from the left boundary  at $10^4$
cm/s, such that a shock with a Mach number of 1.57 was established in
the central gas.  The simulation volume was again two-dimensional,  60
blocks in the $x$ direction by 1 block in the $y$ direction,  and
again 3 additional levels of refinement were allowed, such that the
minimum and maximum zone size in the $x$ direction were $0.0039$ cm
and $0.031$ cm respectively. Finally, along the interface we
initialized  $L=0.01.$

\begin{figure}[ht]
\centerline{\includegraphics[height=10.0cm]{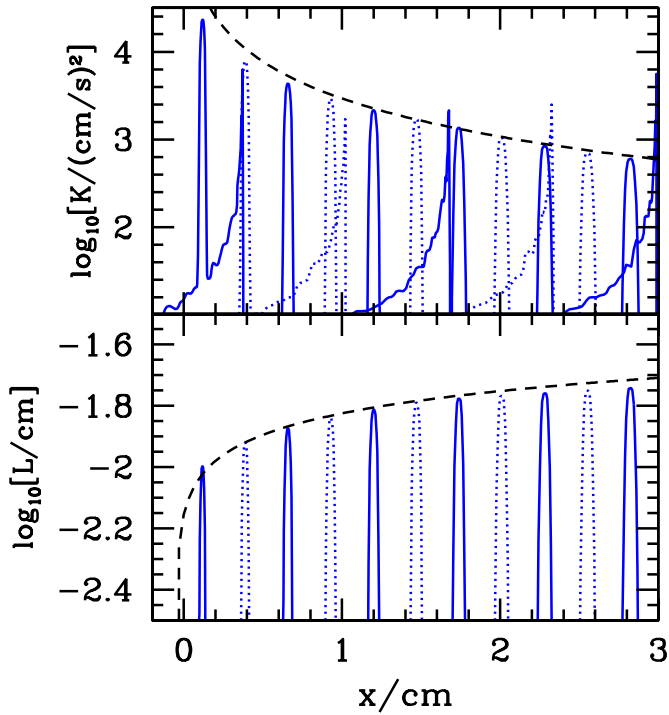}}
\caption{{\em Top:} Profiles of the turbulent kinetic energy per unit
mass at 12 outputs spaced at 25 $\mu$s intervals, beginning at the
moment the shock passes through the interface.  For clarity, 
we alternate outputs between solid lines (25 $\mu$s, 75 $\mu$s, etc.)
and dotted lines (50 $\mu$s, 100 $\mu$, etc.).  The dashed
line shows the analytic solution for the evolution of $K$ at the
interface, eq.\ (\ref{eq:RMK}).  {\em Bottom:} Profiles of the
simulated turbulent length scale at the time steps given in the upper
panels.  The dashed line shows the analytic solution, eq.\
(\ref{eq:RML}).}
\label{fig:RM}
\end{figure}

For this problem the turbulent length scale and kinetic energy are
expected to grow as  \ba L(t) &=& L(0) [t v_{\rm RM}/\theta+1]^\theta,
\label{eq:RML} \\ K(t) &=& K(0) [t v_{\rm RM}/\theta+1]^{2 \theta -2},
\label{eq:RMK} \ea where $\theta = (2 C_D +1.5)^{-1} = 1/4$ where
$v_{\rm RM}$ is the initial post-shock time derivative of $L$ (DT06).
In Figure \ref{fig:RM} we show the $K$ and $L$ profiles that occur
along the density interface as the shock passes through, as compared
to these analytic solutions.  Once we account for the fact that $L$ is
initially compressed as the shock passes through the interface,  eqs.\
(\ref{eq:RML}) and (\ref{eq:RMK}), with $v_{RM} = 1.6 \times 10^4$
cm/s, give an excellent match to the simulation results.  Again, we
repeated this  experiment interchanging the $x$ and $y$, as well as in
the $z$ direction, obtaining identical results.

\section{Cluster Modeling}

Having carried out detailed tests of our subgrid-turbulence model, we
moved on to modify FLASH3 to apply this model to the context of the
AGN heated ICM.  This involved choosing appropriate initial conditions
and metal  injection profiles, modifying the code to account for the
transport and turbulent mixing of  metals, and modeling AGN feedback.

\subsection{Cluster Profile and Metal Injection}

For our overall cluster profile, we adopted the model described in
R07, which was constructed to reproduce the properties of the
brightest X-ray cluster A426 (Perseus) that has been studied
extensively with Chandra and {\sc XMM}-Newton.  In this case,
the electron density $n_{\rm e}$ and temperature $T_{\rm e}$  profiles
are based on the deprojected XMM-Newton data (Churazov \etal 2003;
2004) which are also in broad agreement with the ASCA (Allen \& Fabian
1998), Beppo-Sax (De Grandi \& Molendi 2001; 2002) and Chandra
(Schmidt \etal 2002; Sanders \etal 2004) data. Namely: \be n_{\rm
e}=\frac{4.6\times10^{-2}}{[1+(\frac{r}{57})^2]^{1.8}}+
\frac{4.8\times10^{-3}}{[1+(\frac{r}{200})^2]^{0.87}}~~~{\rm cm}^{-3},
\label{eq:ne}
\ee and \be T_{\rm e}=7\times\frac{[1+(\frac{r}{71})^3]}
{[2.3+(\frac{r}{71})^3]}~~~{\rm keV},
\label{eq:te}
\ee where $r$ is measured in kpc. Furthermore, the hydrogen number
density was  assumed to be related to the electron number density as
$n_{H}=n_{\rm e}/1.2$ according to standard cosmic abundances.  The static, spherically-symmetric
gravitational potential was set such that  the cluster was in
hydrostatic equilibrium.

The metal injection rate in the central galaxy was assumed to be
proportional to the light distribution, and modeled with a Hernquist
profile given by
\begin{equation}
\dot\rho\Metal (r,t) = \frac{\dot M\Metal{}_0} {2\pi} \,  \frac{a}{r}
\, \frac{1}{(r+a)^3},
\label{eq:hernquist}
\end{equation}
with a scale radius of $a= 10$ kpc and a total metal injection rate of
$\dot M\Metal{}_0  =  952\,M_\odot\Myr^{-1}.$ Note that the injection
rate could also have been time-dependent, as used in  Rebusco \etal
(2005), to account for the higher supernova rate in the past and the
evolution of the stellar population (see also Renzini \etal 1993).
However, as we follow the evolution of the cluster for only about
$500$ Myrs,  the metal injection rate does not change significantly
over this time  (Rebusco \etal 2005).

\begin{table*}
\caption{Run Parameters}
\label{tab:runs}
\centering\begin{tabular}{llllllllll} \hline  Run  & $l_{\rm refine}$ &
Maximum    & Effective & Subgrid    & Cooling & Bubble     & Bubble    & Bubble  &
Bubble \\
  Name &                      & Resolution & Grid & Model & &
Type   & Freq. & Radius  & Offset \\
  \hline 5NAES    &  5 & 0.66
kpc &  $1024^3$ & No  & No  &  Evac. & Once      & 11 kpc & 13.2 kpc \\
 5DAES
&  5 & 0.66 kpc & $1024^3$ & Yes & No  &  Evac. & Once      & 11 kpc & 13.2
kpc \\
  3NAES    &  3 & 2.6  kpc & $256^3$ & No  & No  &  Evac. & Once
& 11 kpc & 13.2 kpc \\
  3DAES    &  3 & 2.6  kpc & $256^3$ & Yes & No  &
Evac. & Once      & 11 kpc & 13.2 kpc \\
 4NAES    &  4 & 1.3 kpc & $512^3$
& No  & No &  Evac. & Once      & 11 kpc & 13.2 kpc \\
  4DAES    &
4 & 1.3 kpc  & $512^3$ & Yes & No  &  Evac. & Once      & 11 kpc & 13.2 kpc
\\
  6NAES &  6 & 0.33 kpc & $2048^3$ & No  & No  &  Evac. & Once      & 11
kpc & 13.2 kpc \\
  6DAES    &  6 & 0.33 kpc & $2048^3$ & Yes & No  &  Evac. &
Once      & 11 kpc & 13.2 kpc \\
  5NAES-   &  5 & 0.66 kpc & $1024^3$ &  No  & No
&  Evac. & Once      & 11 kpc & 13.0 kpc \\
  5DAES-   &  5 & 0.66 kpc  & $1024^3$ &  Yes & No &  Evac. & Once      & 11 kpc & 13.0 kpc \\
  5NAES+
&  5 & 0.66 kpc & $1024^3$  & No  & No  &  Evac. & Once      & 11 kpc & 13.4
kpc \\
  5DAES+ &  5 & 0.66 kpc & $1024^3$ & Yes & No  &  Evac. & Once      &
11 kpc & 13.4 kpc \\
  5NAES\_x  &  5 & 0.66 kpc & $1024^3$ & No  & No  &
Evac. & Once     & 11 kpc & 13.0 kpc \\
  5DAES\_x  &  5 & 0.66 kpc & $1024^3$ & 
 Yes & No  & Evac. & Once     & 11 kpc & 13.0 kpc \\
  5NASS    & 5 & 0.66 kpc & $1024^3$ & No  & No   &  
  Sedov  & Once      & 11 kpc & 13.2 kpc \\
5DASS    & 5 & 0.66 kpc & $1024^3$ & Yes & No   &  
Sedov  & Once      & 11 kpc & 13.2 kpc \\
  5NACR    &  5 & 0.66 kpc & $1024^3$ & No  & Yes  &  Evac. & 50
Myr  & 16 kpc & 17.0 kpc\\
  5DACR    &  5 & 0.66 kpc & $1024^3$ & Yes & Yes  &
Evac. & 50 Myr  & 16 kpc & 17.0 kpc \\
  5NCSR    &  5 & 0.66 kpc & $1024^3$ & 
 No  & Yes &  Sedov  & 50 Myr   & 16 kpc & 17.0 kpc \\
  5DCSR    &
5 & 0.66 kpc & $1024^3$ & Yes & Yes  &  Sedov  & 50 Myr   & 16 kpc & 17.0 kpc \\\hline \hline \\
\end{tabular}
\end{table*}

\subsection{Tracing the metals} \label{sec:method_metals}

In order to be able to trace the metal distribution, we utilize a mass
scalar to represent the local metal fraction in each cell, $F \equiv
\rho\Metal/\bar \rho.$ Hence, the quantity $F \bar \rho$ gives the
local metal density, $\rho\Metal$, which has a continuity equation
including the metal source, given by \be \frac{\partial {\bar \rho}
F}{\partial t} + \frac{\partial {\bar \rho} F \tilde u_j}{\partial
x_j} = \frac{\partial}{\partial x_j} \left(  \frac{\mu_t}{N_F}
\frac{\partial F}{\partial x_j}\right)+ \dot\rho\Metal.
\label{eq:continuity_metals}
\ee Furthermore, we assumed that the metal fraction is small at all
times. Hence, we could neglect $\dot\rho\Metal$ as a source term in
the continuity equation for the gas density.

\subsection{Bubble generation} \label{sec:bubble_generation}

Bubbles in the ICM are thought to be inflated by a pair of ambipolar
jets from an AGN in the central galaxy that inject energy into small
regions at their terminal points, which expand until they reach
pressure equilibrium with the surrounding ICM (Blandford \& Rees
1974). The result is a pair of underdense, hot bubbles on opposite
sides of the cluster center.  As in R07 these were modeled using two
methods.

In one method, we injected a total energy of $E\Inj$ over an interval of length $\tau\Inj = 5 \times 10^6$ years into each of two
small spheres of radius $r\Inj$.
The gas inside these spheres was heated and expanded similar to a
Sedov explosion to form a pair of bubbles in a few Myrs, a time much
shorter than the rise time of the generated bubbles.  The parameters
$r\Inj$, $E\Inj$ and $\tau\Inj$ were chosen such that these regions
reached  a radius of $r\Bubble$ and a density contrast of
approximately  $\rho_{\rm b}/\rho_{\rm amb} = 0.05$ as compared to the
surrounding ICM.  However, the dependence of the bubble dynamics on
the density contrast, $\rho_{\rm b}/\rho_{\rm amb}$, is weak provided
that $\rho_{\rm b}/\rho_{\rm amb} \ll 1$ (R07).  In addition to the bubbles,
the explosion sets off shock waves that move through the ICM,
which in fact are the energetically dominant component, as discussed
in more detail below.

In another method, we evacuated the bubble regions by removing gas
while keeping  them in pressure equilibrium with their surroundings.
Inside two spheres of radius $r\Bubble$, we removed the gas at a rate
$\dot\rho$  for a time $\tau\Evac = 5 \times 10^6$ years
that was small compared to the
evolution timescale of the bubbles, but long enough to prevent
numerical problems.  The sink rate $\dot\rho$ was set to decrease the
density inside the bubbles down to a density contrast of $\rho_{\rm
b}/\rho_{\rm amb}$ compared to the surrounding ICM, and in order to
conserve the metal mass, the metal fraction was also updated during
the evacuation.  Since gas was removed from a small volume in the
process of forming the bubble, total mass is not conserved in this
method. However, this does not affect the density profile of the
cluster on the scales that we consider here.  In fact, as illustrated
in  Fig.\ 4 of R07,  the differences between this method and the one
described above are small for the mean-flow variables.  However, the
evacuation method does not cause shocks that move through the ICM,
making it less likely to drive further turbulence  by the
Richtmyer-Meshkov instability.  Thus we used this model to separate
out the effects of buoyancy-driven and shock-driven turbulence on
galaxy clusters, and consider it first in our study.
Note however that in both cases the bubble models are simplified and
do not capture the details of the initial
heating of the ICM by AGN jets, which can also affect stability (\eg
Sternberg \& Soker 2008b)

\subsection{Simulation Parameters}

Recent work has considered  the possibility that AGN bubbles are
stabilized by molecular viscosity  and magnetic fields. Reynolds \etal
(2005) found that even a modest shear viscosity (corresponding to 1/4
of the Spitzer value) can quench fluid instabilities (see also
Ruszkowski \etal 2004).  Smoothed particle hydrodynamic  simulations
of buoyant bubbles in a viscous ICM were also performed by Sijacki \&
Springel (2006).  Furthermore, it has been known for some time that
the ICM hosts significant magnetic fields with typical strengths of 5
$\mu$G  (Carilli \etal 2002).  In many cases, the restoring tension
generated by bending of the field lines may exceed the buoyancy force
driving the RT instability (Chandrasekhar 1961).  De Young (2003)
derived analytic conditions for the stabilization by ICM magnetic
fields, suggesting that observed field strengths might stabilize the
bubble interface in the linear regime, and idea explored
in further detail  by the two-dimensional MHD  calculations in Br\"
uggen \& Kaiser (2001) and recent 3D simulations by Ruszkowski \etal
(2007).  Yet, although viscosity and  magnetic fields can clearly affect
bubble dynamics and turbulence, their impact is still poorly
understood, and we focus here on the basic case of turbulence arising
in an inviscid and unmagnetized fluid.

\begin{figure*}[ht]
\centerline{\includegraphics[height=18.0cm]{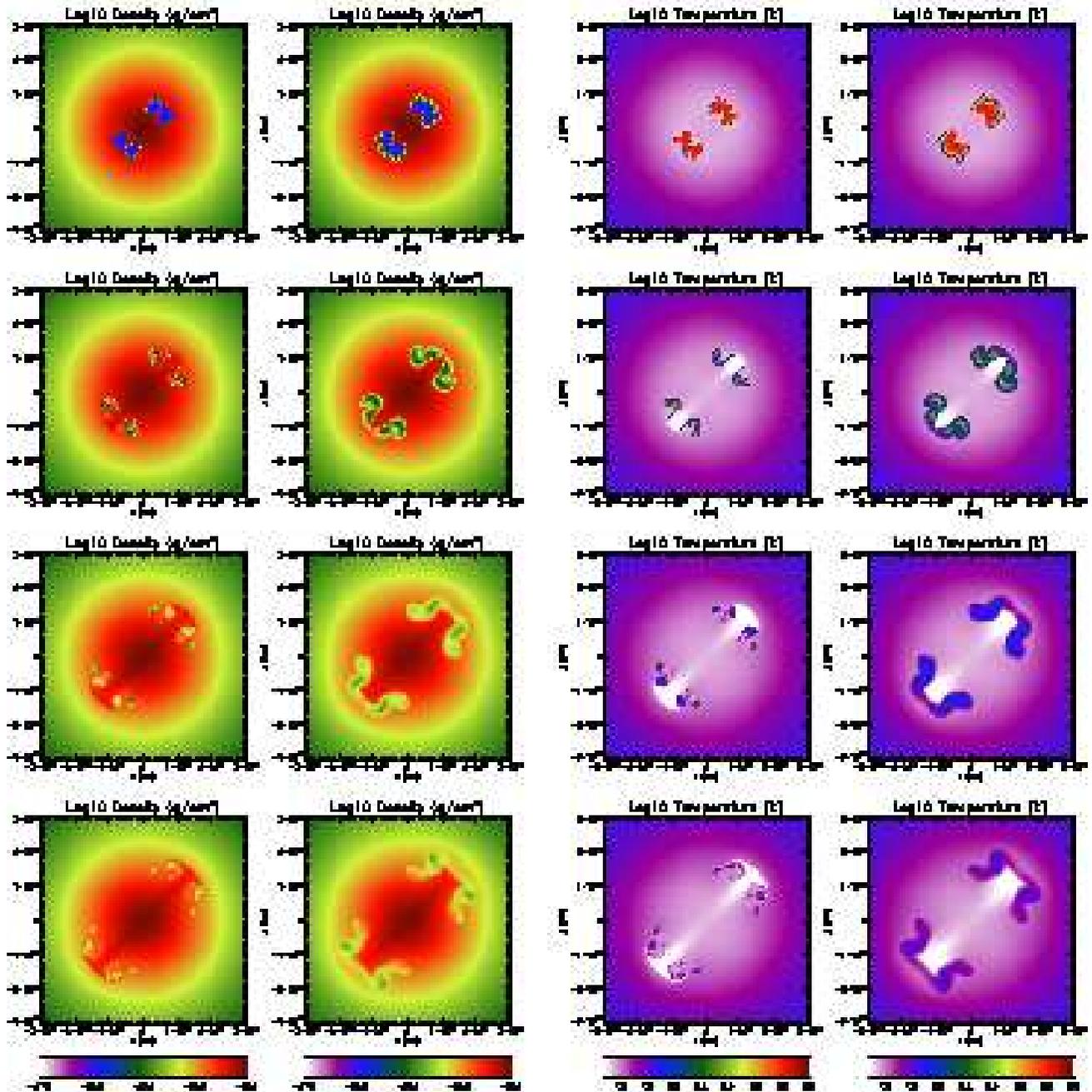}}
\caption{Snapshots of mean-flow quantities in evacuated single-bubble
runs at $t =$ 50 Myrs, 100 Myrs, 150 Myrs, and 200 Myrs (arranged from
top to bottom in each column).   All panels show values at a $z=0$
slice through our simulations and cover the region from $x = -100$ to
100 kpc and $y = -100$ to $100$ kpc.  {\em First Column:} Contours of
$\log \rho$  spanning the range from $\rho =$ $10^{-27}$ g cm$^{-3}$
to $10^{-25}$ g cm$^{-3}$ from the 5NAES pure-hydro run.  {\em Second
Column:} Contours of $\log \rho$ from the 5DAES run with subgrid
turbulence, spanning the same range of densities as in the first
column.  {\em Third Column:}  Contours of $\log T$ from $T= 10^{7.5}$K
to $10^{9}$K from the 5NAES run.  {\em Fourth Column:}   $\log T$
contours from the 5DAES run with the same scale as in column 3.\\}
\label{fig:run1_comparison}
\end{figure*}

All our simulations are performed in a cubic three-dimensional
region 680 kpc on a side, with all reflecting boundaries.  For our
grid, we chose a block size of $8^3$ zones and an  unrefined root grid
with $8^3$ blocks, for a native resolution of 10.6 kpc.   The
refinement criteria are the standard density and pressure criteria,
and we allow for 4  levels of refinement beyond the base grid,
corresponding to  a minimum cell size of 0.66 kpc, and an
effective grid of 1024$^3$ zones.    The parameters
of all the simulations carried out in this study are summarized in
Table \ref{tab:runs}.  We name each run by concatenating the maximum
level of refinement in  FLASH3, the presence of a subgrid-turbulence
model (N  for none or D for DT06), whether the run uses the cooling
model described below (A for adiabatic or C for cooling), the choice
of bubble type (S for Sedov-Type and E for evacuated), and the whether
the runs contain a single pair of bubbles (S) or  periodic (P) AGN
feedback.

\begin{figure*}[t]
\centerline{\includegraphics[height=18.0cm]{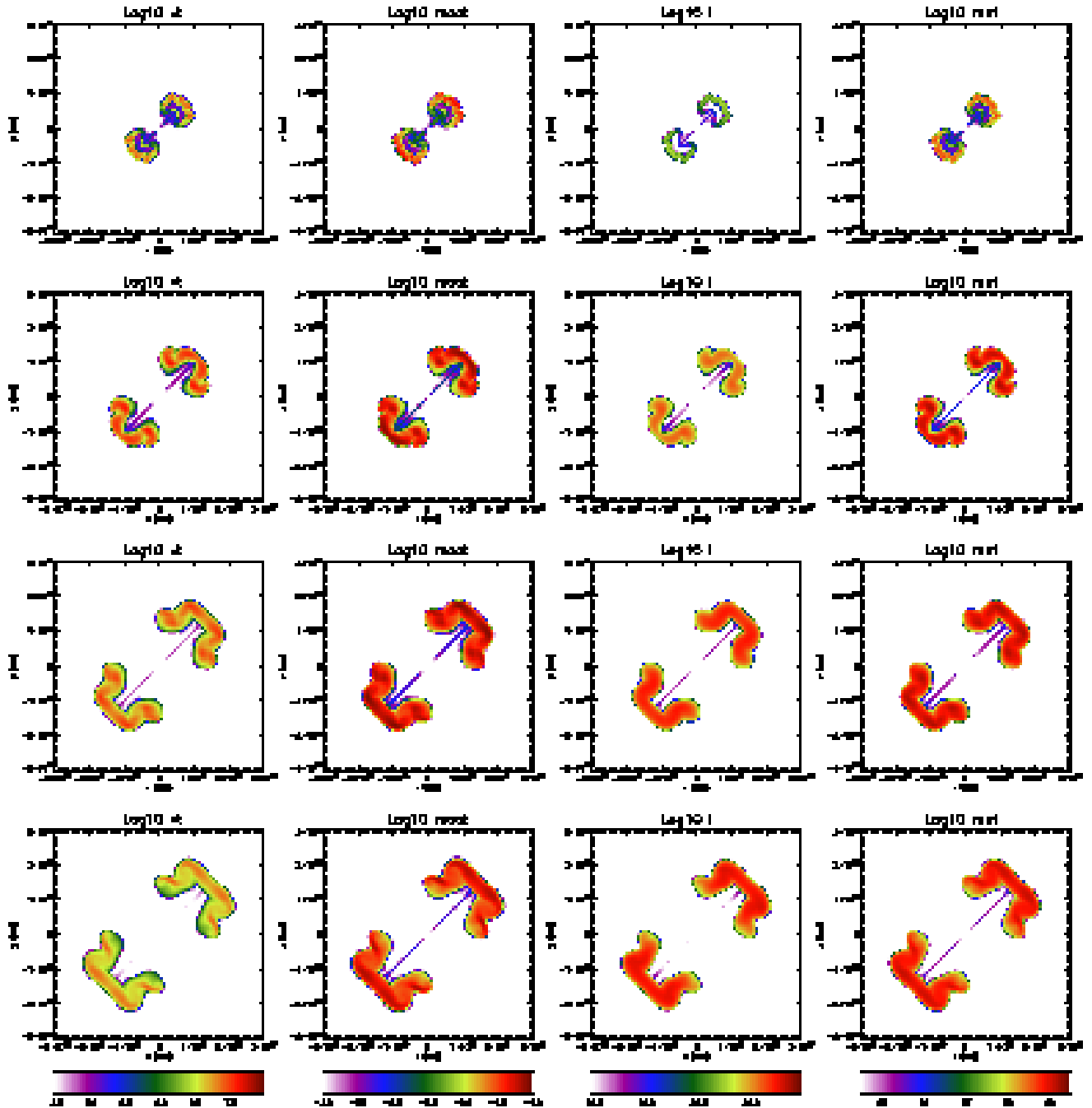}}
\caption{Snapshots of properties of subgrid turbulence in the  5DAES
run  at $t =$ 50 Myrs, 100 Myrs, 150 Myrs, 200 Myrs (from top to
bottom in each column).   As in Figure
\protect\ref{fig:run1_comparison}, all panels show a central 200 kpc
$\times$ 200 kpc $z=0$  slice.  {\em First Column:} Logarithmic
contours of the turbulent velocity labeled in cm s$^{-1}$ 
from $V = \sqrt{2 K} = 0.3$ km
s$^{-1}$ to $V = 300$ km s$^{-1}.$   {\em Second Column:} Logarithmic
contours of the local turbulent Mach number $V/c_s$, from $10^{-4}$ to
$10^{-1}.$  {\em Third Column:} Logarithmic  contours of $L$, labeled
in cm, from
$0.1$ kpc to $10$ kpc.  {\em Fourth Row:}  Logarithmic turbulent
viscosity per unit density $\mu_t /\rho,$ labeled in cm$^2$ s$^{-1}$
from  $10^{-2}$ km/s kpc to $10^3$ km/s kpc.}
\label{fig:run1_turb}
\end{figure*}

\begin{figure*}[ht]
\centerline{\includegraphics[height=12.0cm]{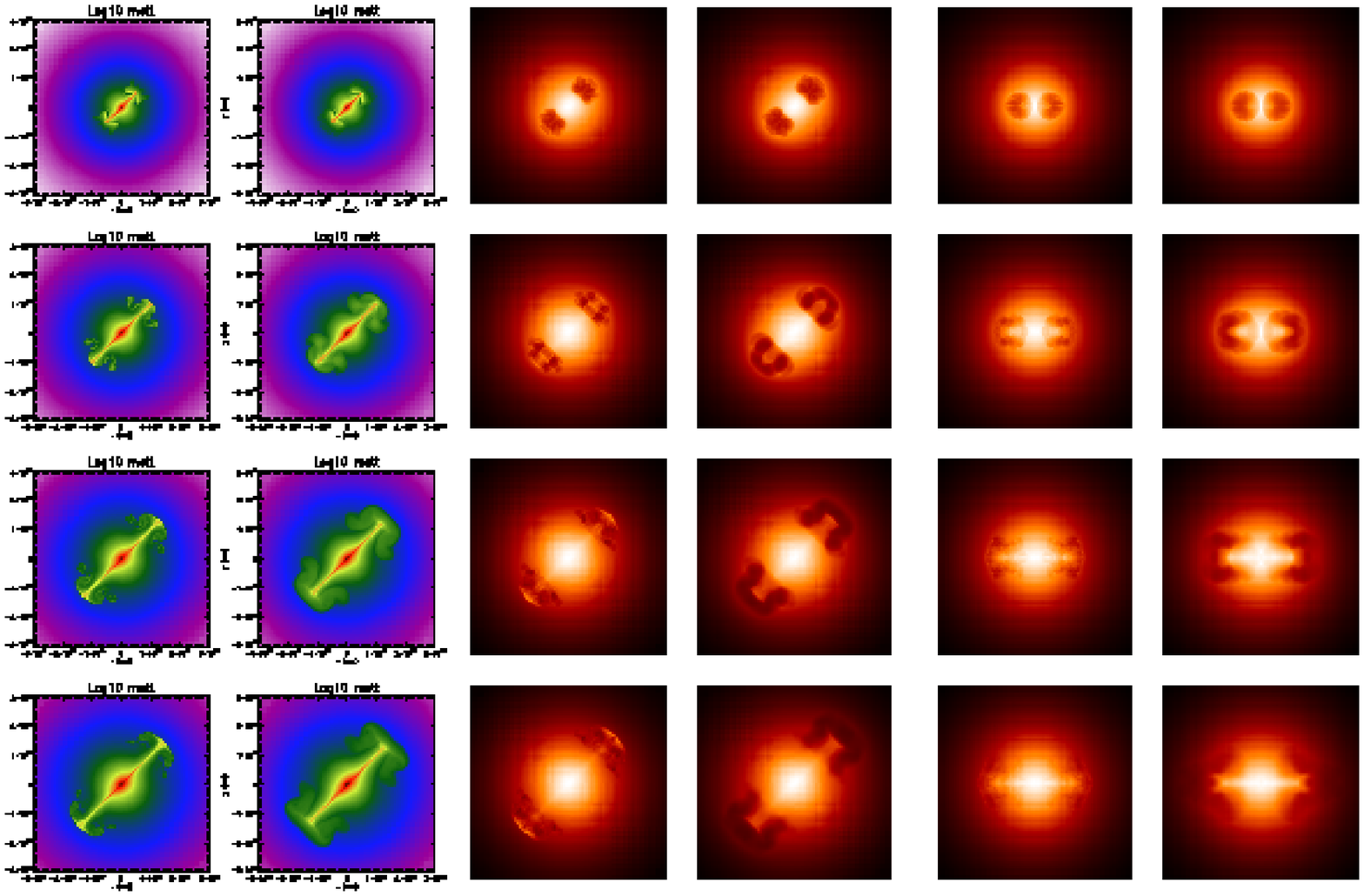}}
\caption{Snapshots of  metal and emission distribution in the central
200 $\times$ 200 kpc slice at $z=0$, at $t=$ 50 Myrs, 100 Myrs, 150
Myrs, and 200 Myrs (from top to bottom in each column) with metallicity
increasing inwards.   {\em First
Column:} Contours of $\log \rho_{\rm metal}$ from the 5NAES  run, at a
fixed but arbitrary scale, with metallicity increasing inwards.  
{\em Second Column:} Contours of $\log
\rho_{\rm metal}$ from the turbulence run, with a scale matching the
first column.  {\em Third Column:} Unsharp X-ray image from the 5NAES
run, projected in the $z$-direction. {\em Fourth Column:}  Unsharp
X-ray image from the 5DAES run, projected in the $z$-direction. {\em
Fifth Column:} Unsharp X-ray image from the 5NAES run, projected in
the $x$-direction.  {\em Sixth Column:} Unsharp X-ray image from the
5DAES run, projected in the $x$-direction.}
\label{fig:run1_observables}
\end{figure*}

\vspace{1.0cm}

\section{Results}

\subsection{Evacuated Bubbles}

As our first case study, we carried out two simulations: an adiabatic
pure-hydro simulation  (5NAES) and an adiabatic simulation with
subgrid turbulence (5DAES).  At the start of  each simulation a
single pair of $r_{\rm bubble} = 11$ kpc bubbles was centered along
the $x=y$ axis at an offset of $R_0 = \pm 13.2$ kpc, and within the
bubbles the gas was evacuated to $\rho_b/\rho_{\rm amb} =0.05$ and
heated to 20 times the temperature of the surrounding medium.
Snapshots of density and temperature from these simulations appear in
Figure \ref{fig:run1_comparison}, spanning times from 50 to 200 Myrs.

The differences between the two runs are dramatic. In the pure-hydro
run the bubbles fragment after rising a single pressure scale
height. The dominant unstable modes that are set by the resolution of
the adaptive grid quickly shred the evacuated regions, drastically
reducing them in size.  If we were able to repeat this simulation with
arbitrarily high resolution, we would find that the bubbles have
developed fingers upon fingers of turbulence, which penetrate deep
into the surrounding medium.  However,  this dispersal is halted by
the finite resolution in the pure-hydro run, which by 100 Myrs has
already significantly underestimated the mixing between the evacuated
region and its surroundings.

On the other hand, the subgrid-turbulence run captures this mixing,
modeling the growth of all the modes that  our computational grid
cannot resolve.  The superposition of these modes then smears
out the interface between the bubble and the ambient medium,
transforming the bubbles into clouds of mixed material, which stay
intact and expand as they rise in the  stratified ICM.
Note that in both runs the bubbles are trailed by streams of colder gas, 
which are lifted from the center and cool adiabatically, consistent with
observations (\eg Simionescu \etal 2007).

To quantify the properties of the turbulent clouds further, in
Figure \ref{fig:run1_turb} we plot slices  from our 5DAES simulation,
focusing on quantities evolved by the subgrid model.  Turbulent
velocities quickly approach a maximum of $\approx 100$ km  s$^{-1},$
roughly $5\%$ of the sound speed within the bubbles and $10\%$ of the
sound speed of the surrounding ICM.  As the hot gas rises at  $\approx
500$ km/s, the turbulent motions are fast enough to mix this material
with the surrounding medium, but much too slow to halt its overall
motion.  In addition to the velocity of turbulent motions, the
efficiency of this mixing process is dependent on the scale of the
turbulent eddies, which is plotted in the third column of Figure 6.
Here we see that $L$ quickly rises to $\approx 10$ kpc, but does not
exceed the overall size of  the clouds, keeping material well-mixed
within these $\approx 30$ kpc regions.  Together $L$ and $V$ act to
produce a typical turbulent viscosity of $\approx 300$ km/s kpc,
which diffuses material  between the clouds and their
surroundings.  Thus rather than being a destructive process,
turbulence acts as an effective {\em mixing} mechanism, which alters
the rising structure, but operates at scales that are too small and
with velocities that are too slow to disrupt it completely.

To study this mixing process further,  we compare the distribution of
ICM metals from our 5DAES (subgrid-turbulence) and 5NAES (pure-hydro)
simulations in the left two columns of  Figure
\ref{fig:run1_observables}.  In both runs, highly-enriched material
from the center of the cluster is carried outward by the rising
bubbles, but this transport is much more effective in the 5DAES run.
In the pure-hydro run metal transport is halted at $\approx 50$ kpc by
bubble disruption, which occurs when RT instabilities shred
the evacuated region into resolution-limited cavities.  In the
subgrid-turbulence run, on the other hand, small-scale fluctuations
act to move metals into the cloud while at the same time
keeping the structure coherent, such that it continues to rise out to
substantially larger radii.

\begin{figure*}[ht]
\centerline{\includegraphics[height=11.0cm]{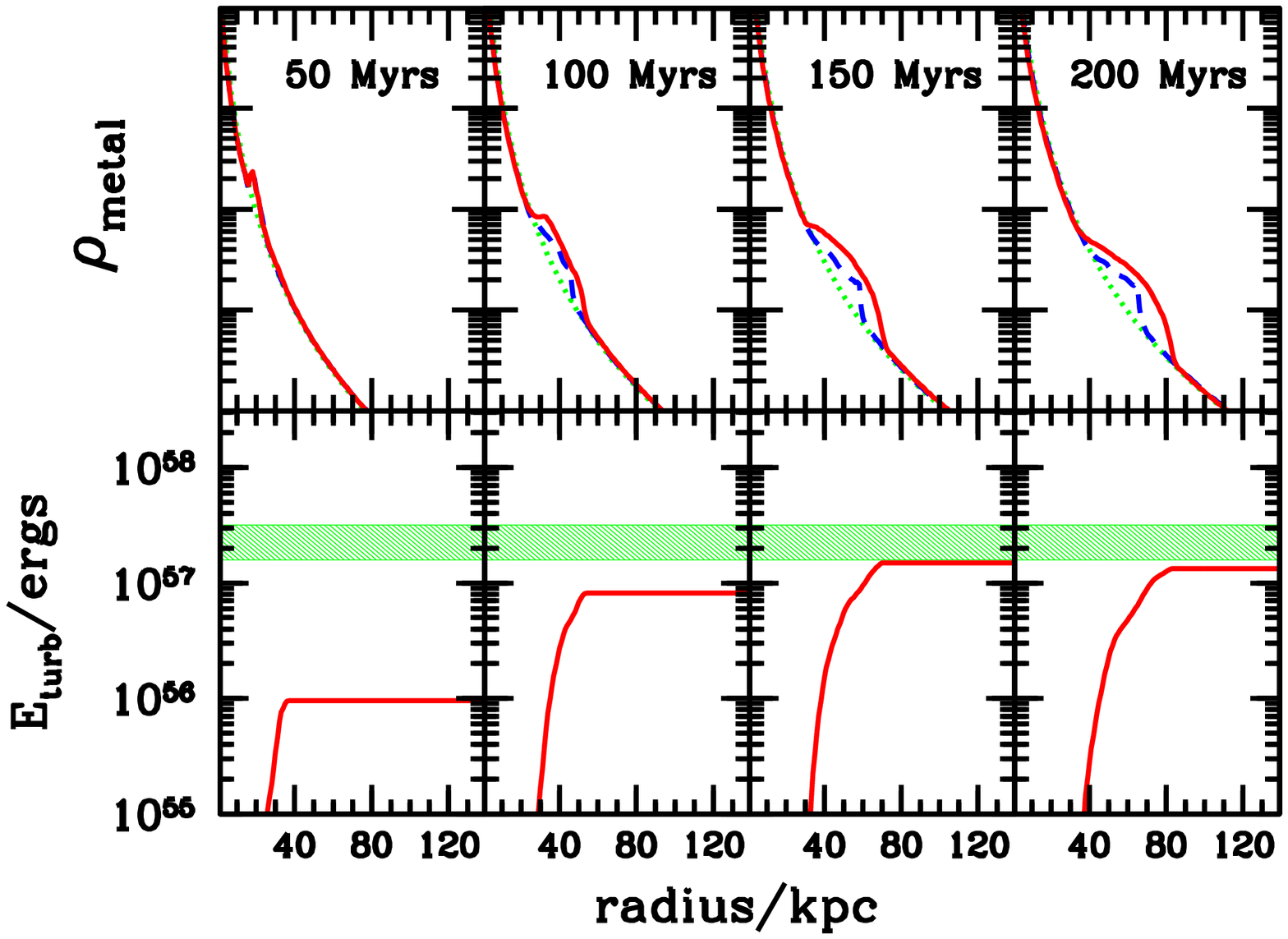}}
\caption{Metal density and total turbulent kinetic energy at times of
50 Myrs, 100 Myrs, and 150 Myrs, 200 Myrs arranged in columns from
left to right.  {\em Top:} Metal density from the evacuated-bubble
subgrid-turbulence run (5DAES, solid lines), and from the evacuated
bubble pure-hydro run (5NAES, dashed lines).  For comparison the
dotted line shows the Hernquist profiles in which metal are ejected
into our simulations.  The logarithmic scale in this row is fixed, but
arbitrary.  {\em Bottom:} Total turbulent kinetic energy contained
within a given radius,  as a function of that radius (solid lines).
The shaded regions are bounded by 1\% to 2\% of $E_{\rm bouyancy}$ as
given by eq.\ (\ref{eq:bouyancy}).}
\label{fig:radialES}
\end{figure*}

We quantify this in the upper row in Figure \ref{fig:radialES}, which
shows the  overall density of metals as a  function of radius in each
of our runs.  By 150 Myrs, the bubbles in the 5NAES run have dissipated,
leaving behind substantial enhancement of metals within 50 kpc, but
having little or no impact outside of this radius.  On the other hand,
clouds in the subgrid-turbulence run continue to rise even at 200
Myrs, leading to enhancements of by over a factor of 5 at 70 kpc.

In the lower panel of Figure \ref{fig:radialES} we plot $E_{\rm
turb}$, the total turbulent kinetic energy within a given radius in
our 5DAES simulation.  This can be directly compared with the total
energy available from the buoyant rise of the bubbles, which can be
simply estimated as (\eg Nulsen \etal 2006)  \be  E_{\rm bouyancy}
\approx \int_{R_0}^\infty V(R) \frac{dp}{dR} dR   \approx \frac{3
p_{\rm evac} V_{\rm evac} }{2},
\label{eq:bouyancy}
\ee where $R$ is the distance of the bubble from the center of the
cluster which is initially at $R_0.$  This gives a value of $\approx
10^{59.2}$ ergs per pair of $r\Bubble = 11$ kpc bubbles.  Comparing
this value to the turbulent kinetic energy in our simulations, we find
that $E_{\rm turb}$ is  only $\approx 1\%$ of this overall energy, thus
we do not expect the energy of the turbulent motion themselves to play
a major role in the heating of the cool-core region in the ICM.
Rather, the key impact of turbulence is to increase the efficiency
with which the thermal energy of the rising cloud is mixed into its
surroundings (\eg Soker 2004; Sternberg \etal 2007; Sternberg \& Soker 2008b).
In this sense it behaves much more like heat conduction 
(\eg Narayan \& Medvedev 2001; Voigt \& Fabian 2004; 
Ruszkowski \etal 2004; Sternberg \etal 2007), than an energy source.
Note however, that our simple turbulence model does not
account for shear-driven effects, and thus may somewhat underestimate
$E_{\rm turb},$ though not by orders of magnitude.


Finally, the physical differences between the two runs lead to
different  observed morphologies.  In the right panels of Figure
\ref{fig:run1_observables} we present approximate X-ray images from
each of the two runs, calculated by projecting the emissivity, computed as
\be 
\epsilon = \Lambda(T) n_e^2,
\label{eq:cool}
\ee 
where we estimate the cooling function, $\Lambda(T),$ which
describes radiative losses from the optically-thin plasma, as in
Sarazin (1986; see also Raymond \etal 1976; Peres \etal 1982).
Furthermore, to draw out structure we create unsharp mask images by
dividing the map by a 30 kpc smoothed version of itself.

These images are comparable to those in the study by Reynolds \etal
(2005), which was carried out with similar resolution, but in a
hydrostatic atmosphere falling off as $\rho(r) \propto [1 +
  (r/r_0)^2]^{-0.75},$ which is significantly less steep than the one
we have prescribed in the central cooling flow region of our
simulation.  Furthermore our times can easily be related to  their
dimensionless units as  $t = \tilde t \times 57 {\rm kpc}/c_s \approx
\tilde t$ 60 Myrs.  Our 5NAES run thus confirms the Reynolds \etal
(2005) result that  pure-hydro inviscid bubbles fall apart within a
dimensionless time of $\tilde t = 3$, or $\approx$ 200 Myrs, although
our steeper radial slope combined with our processing of the image is
able to draw out more structure than visible in their plots.  It is
clear that these collections of fragments look nothing like the smooth
and detached cavities of the type observed in Perseus (\eg Fabian
\etal 2006) and other clusters (\eg  McNamara \etal 2001; Choi \etal
2004;  Reynolds \etal 2008).  On the other hand, in the subgrid-turbulence 
run, the hot clouds lead naturally to coherent holes in the
X-ray distribution not unlike those observed in cool-core clusters. If
anything the bubbles are too large and radially extended as compared
to observations, an issue we take up below, after first addressing the
reliability of our numerical approach.

\begin{figure*}[ht]
\centerline{\includegraphics[height=21.0cm]{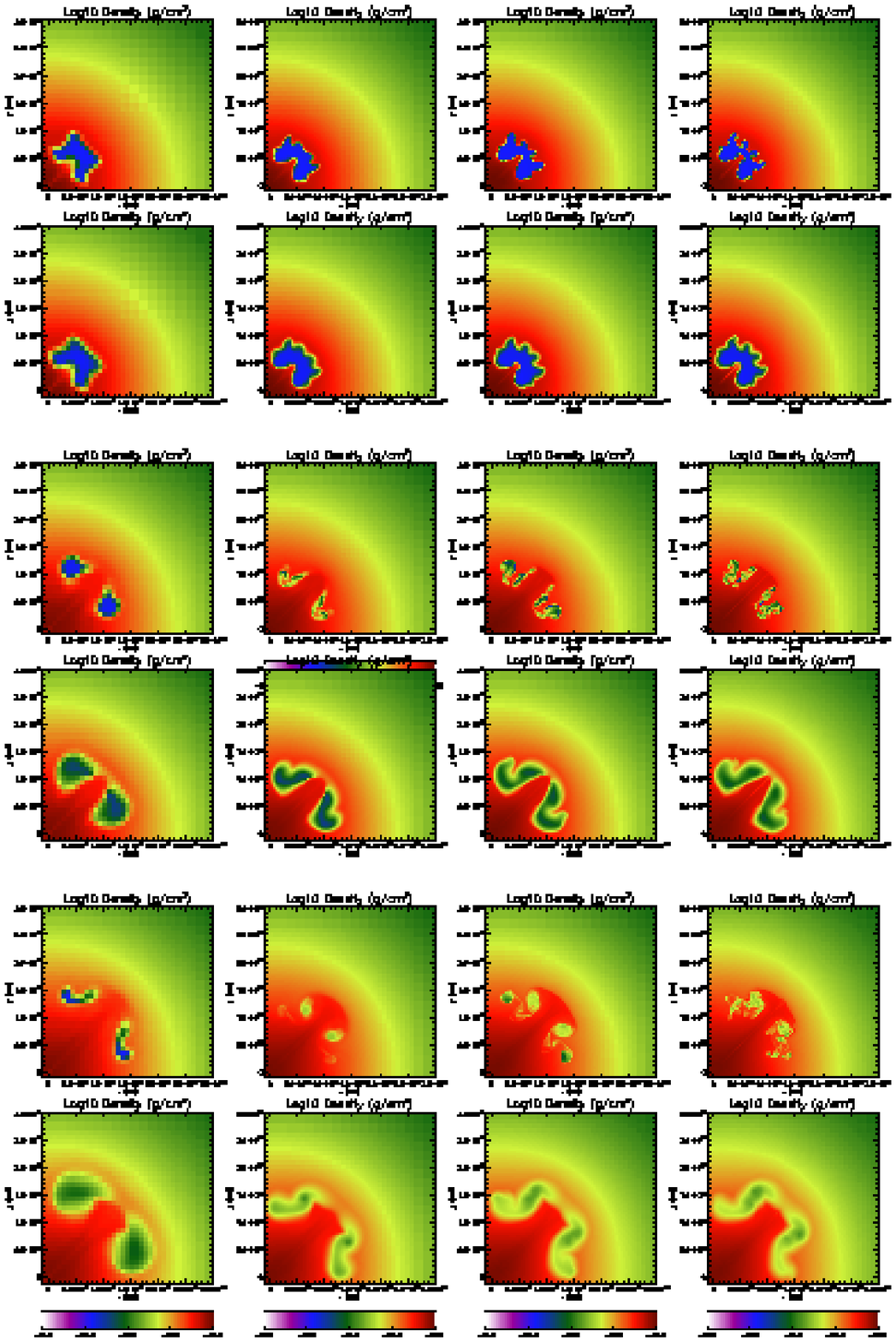}}
\caption{Snapshots of $\log \rho$ in single-bubble runs of varying
resolution.  Plots show a $z=0$ slice of the region from $x = -3$ kpc
to $100$  kpc and and $y =-3$ kpc to $100$ kpc from the pure-hydro
run.  From left to right, each column shows results from simulations
with 3 levels of refinement (2.6 kpc resolution), 4 levels of
refinement (1.3 kpc resolution), the fiducial 5 levels of refinement
(0.66 kpc resolution), and 6 levels of refinement (0.33 kpc
resolution).  {\em Top Row:}    Outputs at 50 Myrs from runs without
the subgrid-turbulence model.  {\em Second Row:} Outputs at 50 Myrs
from runs including subgrid turbulence.  {\em Third Row:}  Outputs at
100 Myrs from runs without subgrid turbulence.  {\em Fourth Row:}
Subgrid-turbulence run outputs at 100 Myrs.  {\em Fifth Row:}  Outputs
at 150 Myrs from runs without subgrid turbulence.  {\em Sixth Row:}
Subgrid-turbulence run outputs at 150 Myrs.}
\label{fig:converge}
\end{figure*}

\begin{figure*}[ht]
\centerline{\includegraphics[height=14.0cm]{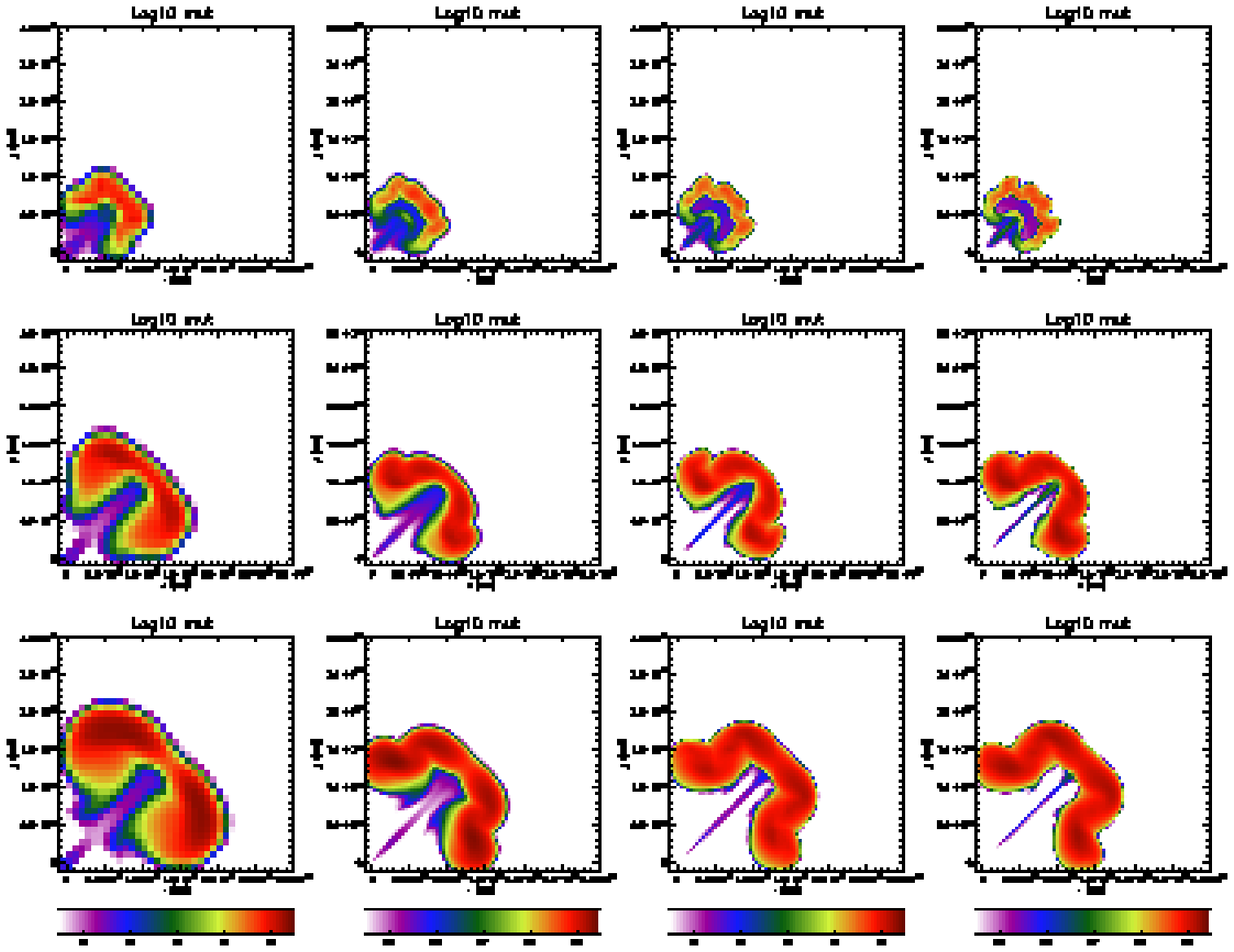}}
\caption{Snapshots of the turbulent viscosity per unit density,
$\log \mu_t/\rho,$ in single-bubble runs of
varying resolution.  As in Figure \protect\ref{fig:converge},  plots
show a $z=0$ slice of the region from $x = -3$ kpc to 100 kpc in the
$x$ and $y$ directions, and from left to right, each column shows
results from simulations with $l_{\rm refine} = 3$ (2.6 kpc
resolution), $l_{\rm refine} = 4$ (1.3 kpc resolution), the fiducial
$l_{\rm refine} = 5$ case (0.66 kpc resolution), and  $l_{\rm
refine} = 6$ (0.33 kpc resolution).   From top to bottom, each row
shows outputs at 50 Myrs, 100 Myrs, and 150 Myrs respectively.  In all
panels $\mu_t /\rho$ is labled in units of cm$^2$ s$^{-1}$ and
goes from from  $10^{-2}$ km/s kpc to $10^3$ km/s
kpc.}
\label{fig:convergeT}
\end{figure*}

\subsection{Dependence on Resolution and Initial Parameters}

An important issue in any simulation is the degree to which its
results are modified when one changes the minimum zone size.  This is
of particular concern  here, because the DT06 model must evolve all
subgrid modes contributing to the RT instability, while still achieving
results that are resolution-independent.   With this in mind, we
carried out a set of convergence tests, repeating our simulations with
and without the subgrid model for cases in which we varied the maximum
level of refinement from $l_{\rm refine} = 3$ (2.6 kpc resolution,
$256^3$ effective grid) up to  $l_{\rm refine} = 6$ (0.33 kpc,
$2048^3$ effective grid).  Snapshots from these simulations are
presented in Figure \ref{fig:converge}, in which we have zoomed into
the area from $x=-3$ kpc to 100 kpc, and $y = -3$ kpc to 100 kpc to
emphasize the regions that are most affected by turbulence.

\begin{figure*}[ht]
\centerline{\includegraphics[height=21.5cm]{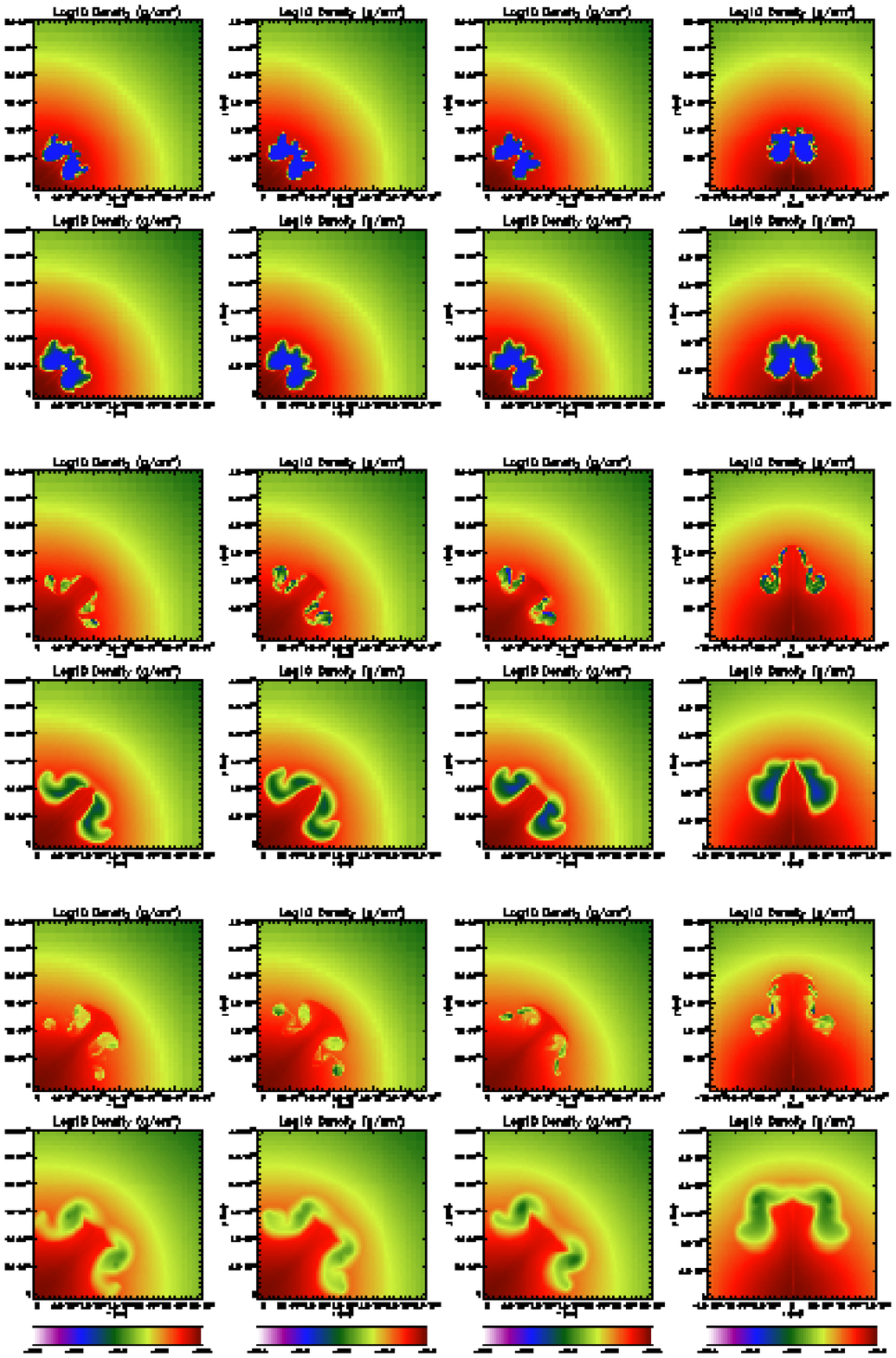}}
\caption{Snapshots of $\log \rho$ in single-bubble runs with slightly
different initial conditions.  As in Figure
\protect\ref{fig:converge}, plots show a $z=0$ slice of the region
from $x = -3$ kpc to $100$  kpc and and $y =-3$ kpc to $100$ kpc.  From
left to right, each row shows results from simulations in which the
bubbles have been initially moved radially by $-1\%$, $0\%$, $1\%$,
and respectively.  Finally, the rightmost row show a run in which the
bubbles are placed along the $y$ axis.  {\em Top Row:}    Outputs at
50 Myrs from runs without the subgrid-turbulence model.  {\em Second
Row:} Outputs at 50 Myrs from runs including subgrid turbulence.  {\em
Third Row:}  Outputs at 100 Myrs from runs without turbulence.  {\em
Fourth Row:} Turbulence run outputs at 100 Myrs.  {\em Fifth Row:}
Outputs at 150 Myrs from runs without turbulence.  {\em Sixth Row:}
Turbulence run outputs at 150 Myrs.}
\label{fig:chaos}
\end{figure*}

Focusing first on the pure-hydro runs at 50 Myrs, one sees a clear
increase in small-scale structure as the runs progress to higher
resolution.   The wavelength of the fastest growing mode from a linear
analysis of the RT instability  is given by (Chandrasekhar 1981) \be
\lambda_{\rm max} = 4\pi (\nu^2 A / g)^{1/3} ,
\label{eq:lambda}
\ee where $A$ is the Atwood number, $g$ is the gravitational
acceleration, and $\nu$ is the viscosity of the fluid, which, in the
pure-hydro run, is given by the effective numerical viscosity.  In the
PPM method used by FLASH, the dissipative processes that act as an
effective $\nu$ are extremely nonlinear.  Not only are the error terms
in this high-order scheme extremely complicated, but they change with
the nature of the flow, as PPM uses several switches that detect
discontinuities (Woodward \& Colella 1984; Porter \& Woodward  1994).
Thus the effective $\nu$ value in our simulations is dependent on the
structures we are trying to resolve, and is best expressed as an
effective Reynolds number $Re  = d \, v/\nu,$ where $d$ and $v$ are
the size and the velocity of the structure of interest.

For our fiducial simulations  (with $l_{\rm refine} = 5$) the
effective number of zones is $1024^3,$ which corresponds to a spatial
resolution of 0.66 kpc.  In the case of large subsonic eddies, not
unlike our bubbles, detailed studies have shown the effective Reynolds
number achievable in PPM simulations is proportional to the number
of grid points across the fluctuation of interest to the power of
$N=3$, where $N$ is the order of the numerical scheme  (Porter \&
Woodward 1994; Balbus \etal 1996; Sytine \etal 2000).  In their study
of viscous bubbles, Reynolds \etal (2005) carried out a series of runs
in which numerical viscosity was explicitly added to the system at
different  levels, and they determined that FLASH3 behaved roughly as
if $Re \approx 2000-5000.$  Using this number as a rough guide in our
fiducial case, which again has resolution similar to that in Reynolds
\etal (2005), this implies that the numerical viscosity is of the
order of  \be  \nu \sim d \, v / {\rm Re} \sim 3 \, {\rm km\, s}^{-1}
\, {\rm kpc} , \ee  where we assumed $d \approx 25 $ kpc and $v
\approx 500$ km/s.  Taking this $\nu$ value, along with $A \approx 1$
and  $g \approx 10^{-8}$ cm s$^{-1}$, this yields a fastest growing
mode wavelength of $\lambda_{\rm max} \approx 2$ kpc, which roughly
corresponds to the scale of the small features seen in the fiducial
pure-hydro  run at 50 Myrs.

From eq.\ (\ref{eq:lambda}), and assuming that viscosity goes as the
number of grid points to the power of 3,  $\lambda_{\rm max} \propto
l_{\rm refine}^{-2}$.  Thus decreasing the resolution to $l_{\rm
refine} = 4$, which corresponds to a spatial resolution of $1.3$ kpc,
moves $\lambda_{\rm max}$ up to  $\approx 8$ kpc, greatly reducing the
perturbations seen at 50 Myrs.  Similarly, choosing $l_{\rm refine} =
3$ moves $\lambda_{\rm max}$ up to the scale of the bubbles, and no
perturbations are visible.   On the other hand, in the $l_{\rm refine}
= 6$ run, $\lambda_{\rm max}$ is reduced almost to the resolution
limit,  leading to perturbations on all simulated scales.   At later
times, the differences between runs persist, such that the bubbles are
shredded into subclumps with sizes that are strongly 
resolution-dependent.  Note however, that the fastest growing 
mode as given by
eq.\ (\ref{eq:lambda}), need not be precisely the mode that dominates
the late-time nonlinear evolution of the subclumps,  as the growth of
perturbations at $\lambda_{\rm max}$ may saturate.

\begin{figure*}
\centerline{\includegraphics[height=12.5cm]{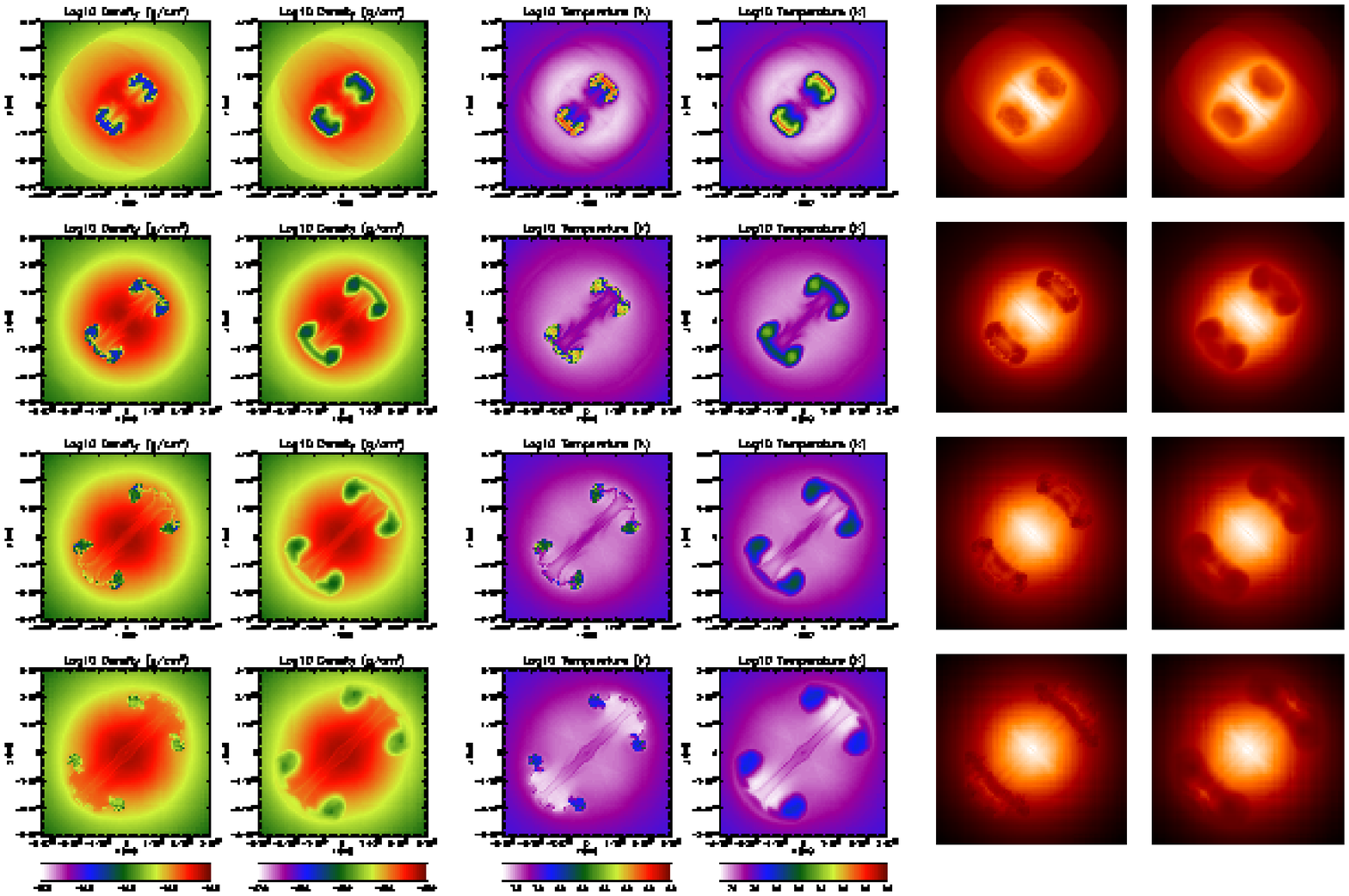}}
\caption{Snapshots of mean-flow quantities in Sedov single-bubble
runs at 50 Myrs, 100 Myrs, 150 Myrs, 200 Myrs (arranged from top to
bottom in each column).   All panels show values at a $z=0,$ slice
through our simulations and cover the region from $x = -100$ to 100
kpc and $y = -100$ to $100$ kpc, with scales chosen as in previous
plots. {\em First Column:} Contours of $\log \rho$ from the run
without subgrid turbulence, spanning the range from $\rho =$
$10^{-27}$ g cm$^{-3}$ to $10^{-25}$ g cm$^{-3}.$ {\em Second Column:}
Contours of $\log \rho$ from a run with subgrid turbulence, and
spanning the same range of densities as in the upper panels.  {\em
Third Column:}  Contours of  $\log T$ from $T= 10^{7.5}$K to
$10^{9}$K, from the run without subgrid turbulence.  {\em Fourth
Column:}   $\log T$ contours from the turbulence run, again with the
same dynamical range.  {\em Fifth Column:} Unsharp X-ray image from
the 5NASS run, projected in the $z$-direction.  {\em Sixth Column:}
Unsharp X-ray image from the 5DASS run, projected in the $z$-direction.}
\label{fig:shockmean}
\end{figure*}

Moving to  the subgrid model runs, we find no such dependence on
resolution.  In this case, the code carries two additional quantities,
namely the turbulent length scale, $L$, and kinetic energy, $K$. It
evolves $L$ and $K$ to reproduce the analytic growth of the RT and RM
perturbations in the limit in which the molecular viscosity is small
and does not play a  role on the scales of interest.   The growth of
small-scale perturbations is then completely captured by the evolution
of $L$ and $K,$  which  are used to construct a turbulent viscosity
that is imposed on the mean-field variables explicitly.  The role of
the numerical viscosity in eq.\ (\ref{eq:lambda}) is then played by
$\mu_t/\rho,$ which  grows to a value of 300 km/s kpc, larger than the
numerical viscosity in all of the runs. Thus rather than breaking up
into a collection of resolution-dependent subgroups, the mean-field
density distributions remains coherent, while at the same time the
simulation is able to maintain the information necessary to calculate
the mixing of the turbulent material with the surrounding medium.

Furthermore,  the evolution of the turbulent viscosity is determined
by the  overall properties of the flow, and is largely independent of
resolution.  A test of this convergence is shown in Figure
\ref{fig:convergeT}, which  gives snapshots of  $\log \mu_t/\rho$ in
each of our single-bubble runs with varying resolution. Here we see
that $\mu_t/\rho,$ whose evolution is determined by the gravitational
acceleration and the Atwood number, evolves to $\approx  300$ km/s kpc
regardless of our choice of $l_{\rm refine}.$  Thus the overall
structure of evolving  bubbles is quite similar across runs.

The subgrid-turbulence model also greatly reduces
the sensitivity of our results on the detailed choice of initial
conditions.  Figure \ref{fig:chaos} shows the results of pure-hydro
and subgrid simulations in which we have taken the initial offset of
the bubbles to be $R_0 = 13.0$, $13.2$, and  $13.4$ kpc, as well as a
pair of runs in which we have taken $R_0 = 13.0$ kpc, but  oriented
along the $y$-axis.  In the pure-hydro case, our results show a strong
sensitivity to these $\pm 2\%$ changes in initial position.  This is
because the code is attempting to sample an underlying turbulent
density field with resolution-dependent subclumps, whose  locations
vary significantly depending on the nonlinear, position- and
velocity-dependent effective numerical viscosity.  In the subgrid runs
however, the presence of the turbulent viscosity leads to a smoothing
over the unresolved turbulent density distribution, thus capturing the
full range of positions over which fully-resolved turbulent material
would be spread.

\subsection{Sedov Bubbles}

Next we turned our attention to the more computationally-demanding
case in which the bubbles are modeled as initially overpressured
spheres.  In this case we raised the internal temperature such that
each bubble expanded to a density of $\rho_b/\rho_{\rm amb} = 0.05$
before reaching pressure equilibrium with a bubble size equal to the
one in the evacuated model.  For a $\gamma = 5/3$ gas, the energy
input during this expansion can be simply  calculated as 
\be 
E_{\rm expand} = \int_{V_{\rm init}}^{V_{\rm evac}} dV p
= \frac{3 p_{\rm final} V_{\rm evac}}{2}  \left[ \left( \frac{V_{\rm
  evac}}{V_{\rm init}} \right)^{2/3} -1 \right],
\label{eq:expand}
\ee where $p_{\rm evac}$ and $V_{\rm evac}$ are the pressure and
volume at the evacuated stage at the end of the expansion and $V_{\rm init}$
is the initial volume.   Although the bubble expansion occurs at
approximately the sound speed of the internal material, this velocity
is well above to the sound speed of the exterior gas.  This energy is
then deposited into shocks which act both to directly raise the
temperature in the surrounding medium, as well as increase the
turbulent kinetic energy through the Richtmyer-Meshkov instability as
the shocks meet acoustic discontinuities.
This energy input can be contrasted to the energy
input from the buoyant motions of the bubbles studied in \S5.1, as
given by eq.\ (\ref{eq:bouyancy}).  For our choice of
$\rho_b/\rho_{\rm amb} = 0.05$,   $\left(V_{\rm evac}/V_{\rm init}
\right)^{2/3} -1  = 6.35$.    Thus $E_{\rm expand}/E_{\rm buoyancy}
\approx   6$, such that overpressured bubbles should have a
significantly greater impact on the cluster than the evacuated model.

The impact of this  energy is seen in the outer oval-shaped shock
visible in the 50 Myr  plots of density, temperature, and simulated
X-rays shown in Figure \ref{fig:shockmean}.  Trailing behind this
discontinuity is a second density and temperature jump,  which is due
to initially inward-facing shocks that reflect off each other and
move into the surrounding medium.   The motion of these reflected
shocks through the rising bubbles has a noticeable effect on the bubbles'
morphology, substantially compressing them in the radial direction.
At later times in the pure-hydro run, the  result is a flattened
distribution of distinct cavities, which separate and shred as the
bubbles move outwards, much as they did in the 5NAES simulation.

In the subgrid-turbulence case, however, the result is a flattened but
coherent cloud  which  moves  outwards, mixing with the surrounding
medium while remaining intact.  This distinctive shape immediately
brings to mind familiar images of  the ``mushroom cloud'' structure
that results from a large airburst in the atmosphere of Earth.  Indeed,
the language used to describe such clouds, such as  a ``toroidal fireball''
perched  atop a somewhat cooler updrafting ``stem,'' (\eg Glasstone \&
Dolan 1977), translates naturally to describe the structures we see in
our simulations.  In fact, the physics of these two phenomena are
deeply connected.  In both cases an unstable bubble is rising
upward in a hydrostatic medium. In  both cases this bubble  is neither
stabilized by magnetic fields nor through viscosity.   And in both
cases the hot gas is well-mixed with the  surrounding medium as it
moves through it, yet it manages to retain its shape.  Furthermore,
as shown in the rightmost panels of Figure \ref{fig:shockmean}, these
radially-flattened clouds lead naturally to X-ray ``holes'' of
precisely the type observed in  cool-core clusters, without requiring
additional physics.

\begin{figure*}[ht]
\centerline{\includegraphics[height=18.0cm]{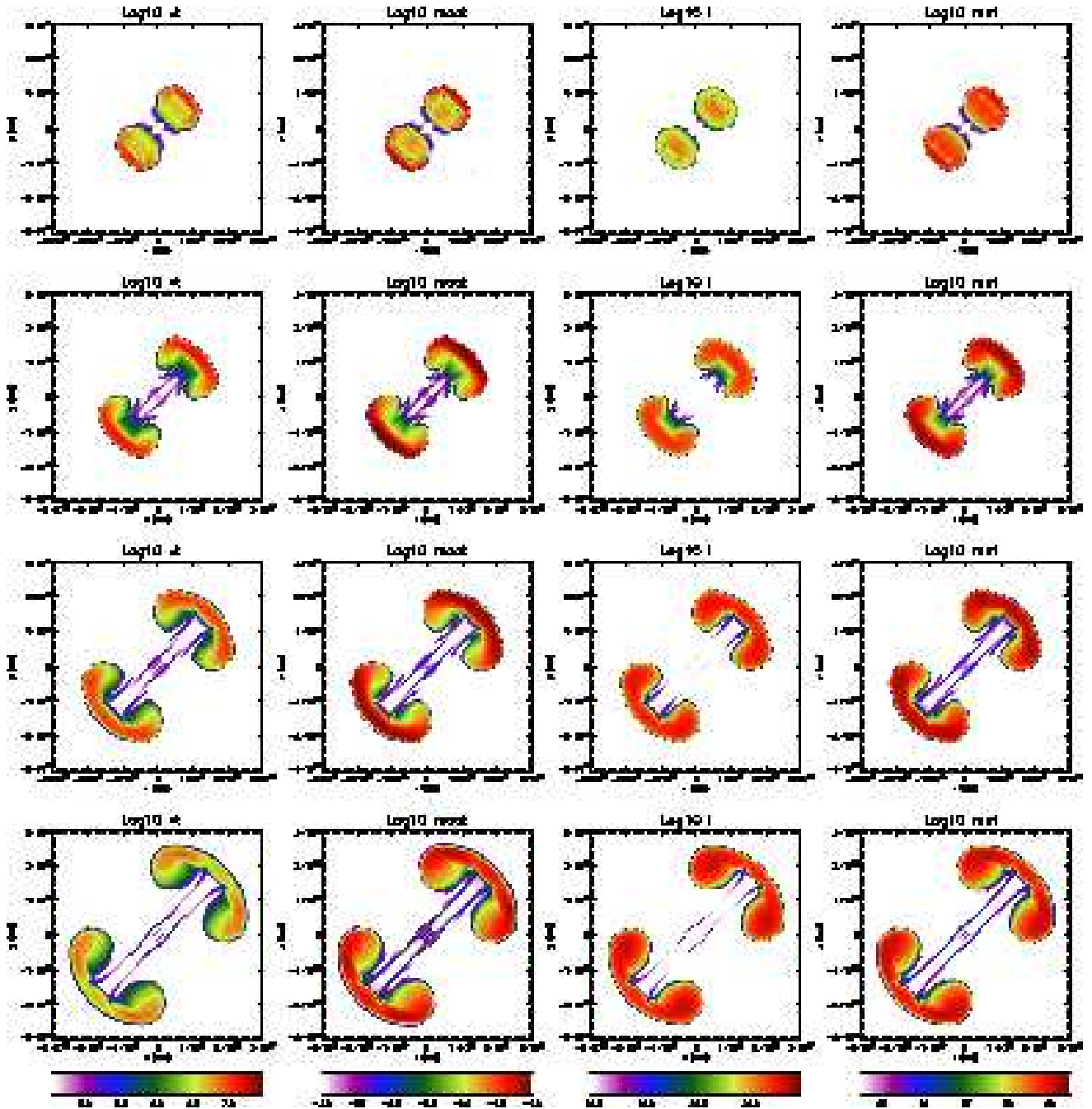}}
\caption{Snapshots of properties of subgrid turbulence in the Sedov
single-bubble (5DASS) run  at $t =$ 50 Myrs, 100 Myrs, 150 Myrs, 200
Myrs (from top to bottom in each column).   All panels show a central
$z=0$,  200 kpc $\times$ 200 kpc slice.  {\em First Column:}
Logarithmic contours of the turbulent velocity from  0.1 km s$^{-1}$
to 100 km s$^{-1}.$   {\em Second Column:} Logarithmic contours of the
local turbulent Mach number from $10^{-4}$ to $10^{-1}$  {\em Third
Column:} Logarithmic  contours of $L$ from $0.1$ kpc to $10$ kpc.
{\em Fourth Row:}  Logarithmic turbulent viscosity per unit density
from  $10^{-2}$ km/s kpc to $10^3$ km/s kpc.}
\label{fig:shockturb}
\end{figure*}

In Figure \ref{fig:shockturb} we show the evolution of the turbulence
variables in the 5DASS run.  In general, the turbulent velocities and
length scales  are systematically enhanced with respect to the
evacuated-bubble run, due to the effect of the RM instability.  In
this case, turbulent velocities can  approach 300 km/s, 50\% greater
than 5DAES run, but still well below the $\approx 500$ km/s radial
velocity of the cloud.   Similarly, the turbulent length scales grow
faster at early times and remain somewhat higher at later times, as
compared to the evacuated case, but they never exceed the overall
scale of the rising cloud.  Together $V$ and $L$ act to generate a
typical turbulent viscosity of $10^3$ km/s kpc,  a substantial  value
that is likely to significantly influence the overall evolution of the
intracluster medium.

\begin{figure*}
\centerline{\includegraphics[height=8.0cm]{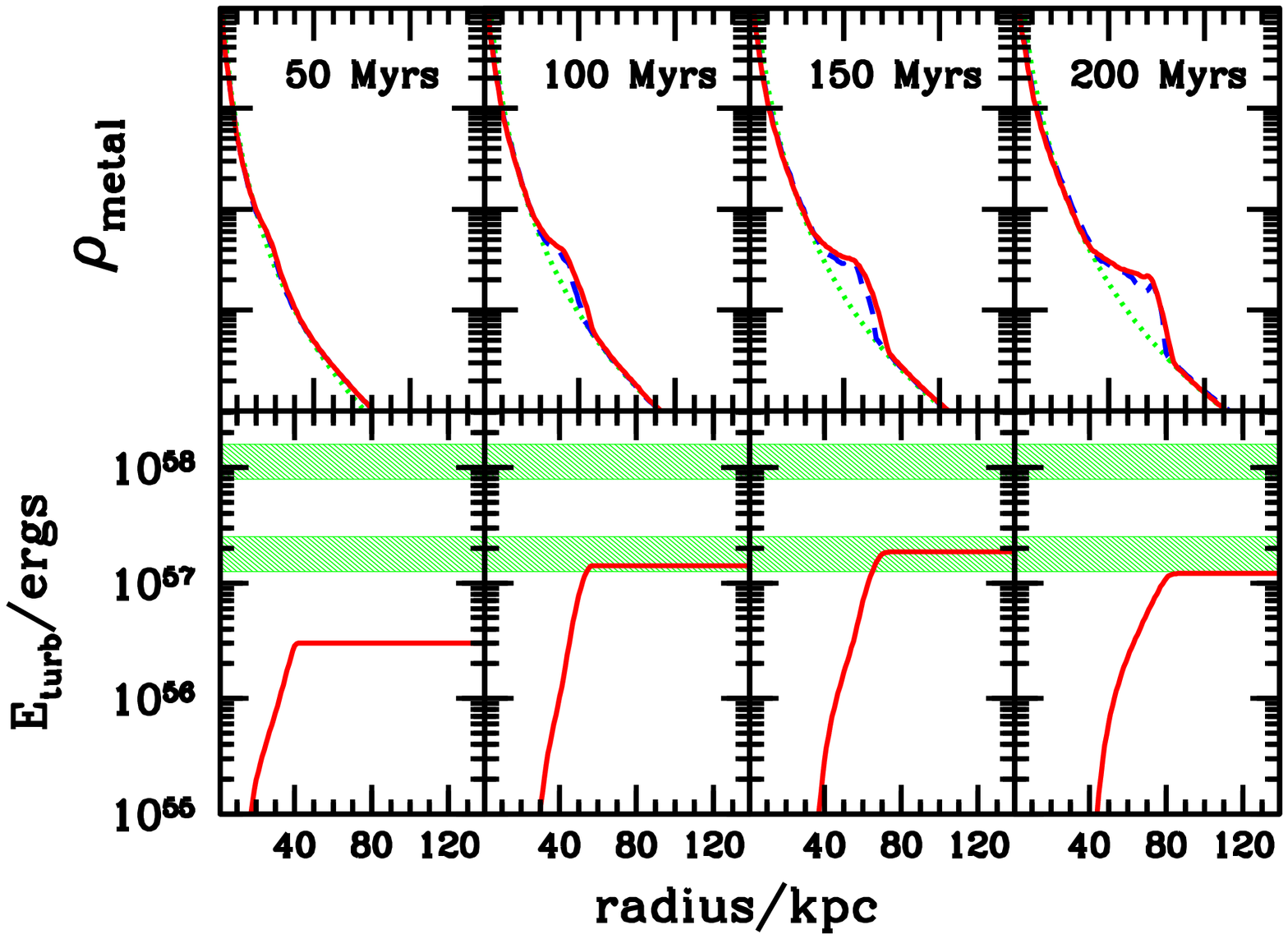}}
\caption{Metal density and turbulent kinetic energy (bottom row) from
the Sedov single-bubbles runs at times 50 Myrs, 100 Myrs, and
150 Myrs, 200 Myrs arranged in columns from left to right.   {\em
Top:} As in Figure \ref{fig:radialES}, the solid lines are drawn from
the  subgrid-turbulence run (5DASS), the dashed lines are from the
pure-hydro run (5NASS), and the dotted line is the profile in which
metals are injected.  {\em Bottom:} Total turbulent kinetic energy
within a given radius in our simulations (solid lines)  contrasted
with 1-2\% of $E_{\rm bouyancy}$ and 1-2\% of $E_{\rm expand} + E_{\rm
bouyancy},$ represented by the lower and upper shaded regions,
respectively.}
\label{fig:radialSS}
\end{figure*}
\begin{figure*}
\centerline{\includegraphics[height=12.1cm]{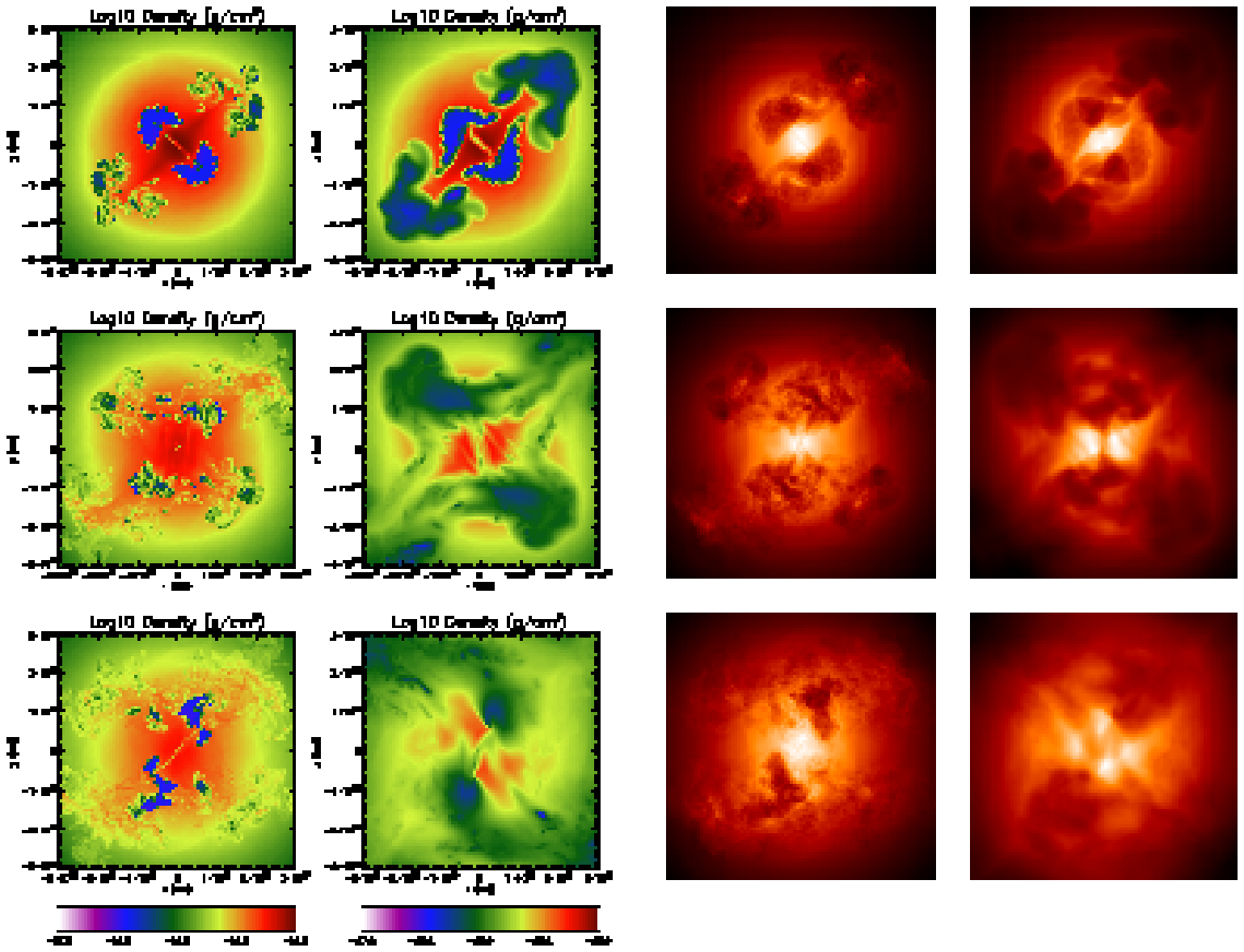}}
\caption{Snapshots of mean-flow quantities in the periodic evacuated-bubble 
runs at 150 Myrs, 300 Myrs, and 450 Myrs (arranged from top to
bottom in each column).   All panels show values at a $z=0$ slice
through our simulations and cover the region from $x = -100$ to 100
kpc and $y = -100$ to $100$ kpc, with scales chosen as in previous
plots. {\em First Column:} Contours of $\log \rho$ from the run
without subgrid turbulence, spanning the range from $\rho =$
$10^{-27}$ g cm$^{-3}$ to $10^{-25}$ g cm$^{-3}.$ {\em Second Column:}
Contours of $\log \rho$ from a run with subgrid turbulence, and
spanning the same range of densities as in the upper panels.{\em Third
Column:} Unsharp mask X-ray image from the  5NCER run, projected in
the $z$-direction. {\em Fourth Column:}  Unsharp mask X-ray image from
the  5DCER run, projected in the $z$-direction.}
\label{fig:periodicmean}
\end{figure*}

In Figure \ref{fig:radialSS} we quantify the radial distribution of
metals and turbulent energy in our Sedov-bubble simulations.   In
this case the differences in $\rho_{\rm metal}$ between the pure-hydro
and subgrid-turbulence runs are much less dramatic than they were in
the evacuated-bubble models.  The overall shock heating of the central
region drives metals outwards to $\approx 80$ kpc in 200 Myrs,
regardless of whether they are transported in well-mixed clouds or
fragmented cavities.  A comparison of $E_{\rm turb}$ to the energy
from the buoyant bubbles shows that while the Richtmyer-Meshkov
instability increases turbulence somewhat from the evacuated case,
the magnitude of this change is much smaller than the factor of $\approx 7$
increase in available energy.  Most of the excess  energy instead
goes directly into post-shock heating or establishing large-scale
motions.  Thus even more than in the evacuated case, turbulence plays
a role primarily in determining mixing and structure, rather than in
providing a source of energy to balance cooling.

\subsection{Periodic Bubbles}

Finally, we carried out a set of
simulations to study the impact of our subgrid-turbulence models on
the evolution of cool-core clusters on longer time scales.  This
involved two major changes from our single-bubble simulations.

First, we implemented a model in which AGN-driven bubbles occur
periodically, again following an approach developed in R07.    As in
our previous runs, we generated bubbles in pairs, but now at regular
intervals  of $\tau\Bubble = 50$ Myrs.  Furthermore, we rotated the
position of the center of the bubble pairs by 90 degrees around the
$y$-axis.  Thus, the first pair of bubbles was centered  at $z=0$ and
$x=y= \pm R_0/\sqrt{2}$, the second pair of bubbles at $x=0$ and
$y=z= \pm R_0/\sqrt{2},$  the third at $z=0$ at $x= -y= \pm R_0/\sqrt{2}$, 
etc., cycling back to the original position every 200 Myrs.


The second major change to our simulations was the addition of
cooling, which was calculated in the optically-thin limit from the
same emissivity as used to construct our  X-ray images, namely
$\epsilon = \Lambda(T) n_e^2$, where the cooling function $\Lambda(T)$
was taken from Sarazin (1986).  However, in order to avoid drastic
cooling within a single timestep  we did not cool cells in which the
density was above $5 \times 10^{-25}$ g cm$^{-3}$ or in which the
temperature was below $5 \times 10^{6}$K.  This placed an absolute
entropy floor in our simulations at 5 keV cm$^2$.

With this cooling in place, we found that our simulation was initially
radiating  at $\approx 4 \times 10^{44}$ ergs per second or   $\approx
1.2\times 10^{60}$ ergs per 100 Myrs.   Thus we chose to scale up our
bubbles to a radius of $r\Bubble = 16$ kpc such that  $E_{\rm
buoyancy} \approx 5 \times 10^{59}$ ergs, and the energy input from buoyant
heating roughly balanced cooling.  As before, the parameters for our
runs are summarized in Table \ref{tab:runs}.

Figure \ref{fig:periodicmean} shows snapshots of these simulations in
the evacuated-bubble case at times of 150, 300, and 450 Myrs.  Like
the single-bubbles runs, the pure-hydro simulation (5NCER) shows
significantly more fragmentation than the  subgrid turbulence run
(5DCER), particularly towards the outskirts of the cluster.   However,
unlike  the single-bubble runs, the properties of the two simulations
continue to diverge with time at all radii.   This is because
turbulence driven by previous feedback events remains behind,
enhancing the mixing of subsequent bubbles.

Thus, at late times, the X-ray maps from the subgrid-turbulence run
contain two prominent holes, while the distribution in the
pure-hydro run is much more fragmented.  Note also that by 450 Myrs,
errors introduced by  the split-hydro solver employed by FLASH3 have
grown to the point that  substantial assymetries are seen in the
pure-hydro runs.   As we saw in Figure \ref{fig:chaos} this dependence
on small fluctuations is reduced in the subgrid turbulence run,
as the turbulence viscosity smoothes over what are essentially
different realizations of the same turbulent density field,  a field
that has been unnaturally quantized into resolution-dependent
subclumps  in the pure-hydro runs.

\begin{figure}[ht]
\centerline{\includegraphics[height=9.0cm]{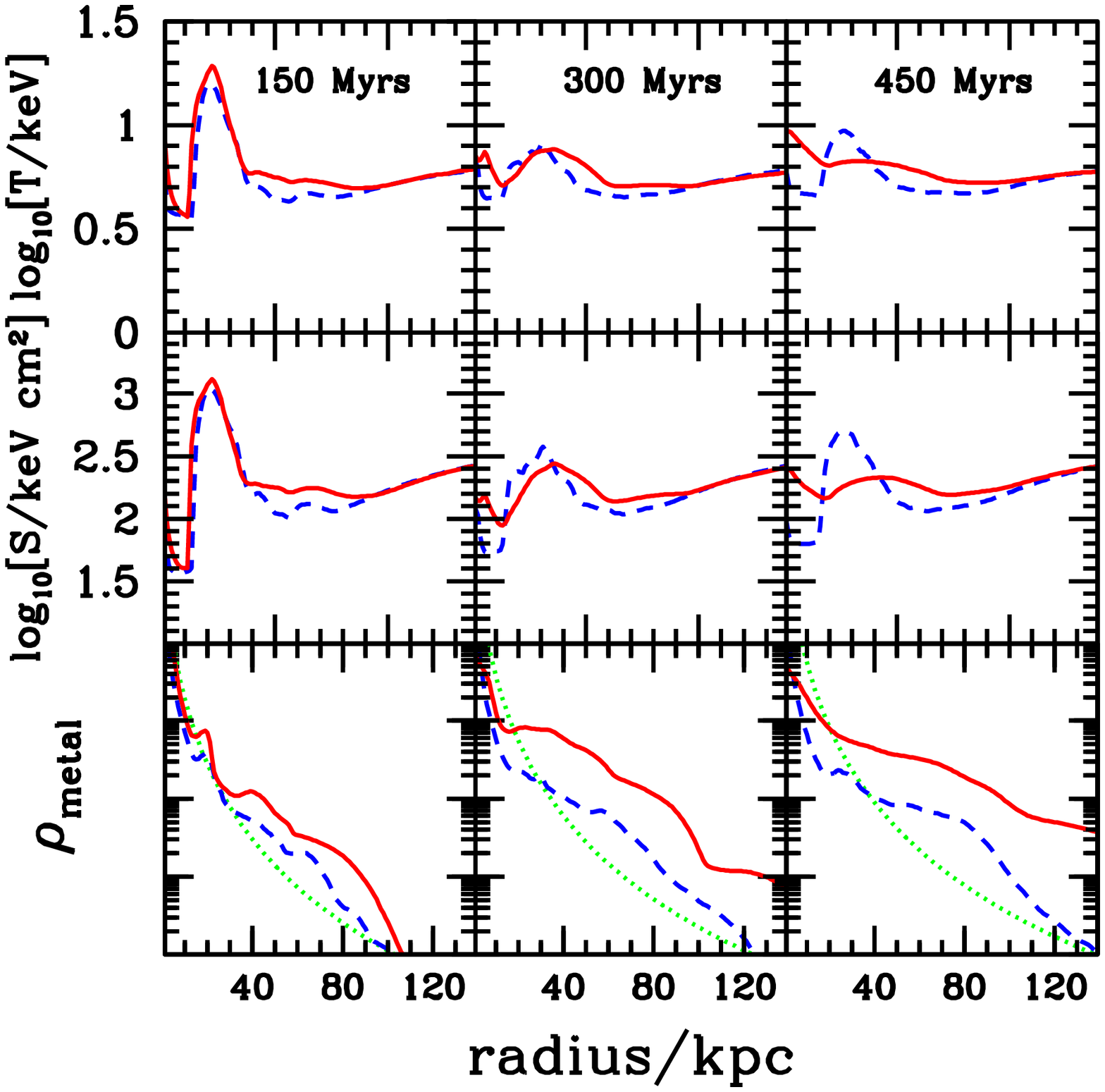}}
\caption{Radial profile of mean-flow quantities in evacuated periodic
bubble runs at times of 150 Myrs, 300 Myrs, and 450 Myrs, arranged in
columns from left to right.  From top to bottom, the rows show the
average temperature, entropy, and metal density.  In all panels the
solid lines are taken from the subgrid-turbulence run (5DCER) and the
dashed lines are taken from the  pure-hydro run (5NCER).   The dotted
lines in the bottom panels show the metal profile that would be found
if there were no resolved or subgrid velocities in the simulations.}
\label{fig:radialER}
\end{figure}

Figure \ref{fig:radialER} shows the evolution of the radial profiles
of mean-field quantities in our simulations.  Focusing first on
temperature, clear differences are apparent between the two runs, and
these differences grow stronger with time.    Within the central 20
kpc the temperature in the subgrid-turbulence  run is roughly a factor
of two greater than in the pure-hydro run, and a similar difference is
apparent in the radial entropy profile.  As suggested by the
single-bubble  runs, and as quantified for the 5DCER run below, these
differences are {\em not} due to additional energy input  from
dissipated turbulent kinetic energy.  Rather, the primary role of
turbulence is to improve the mixing of the thermal energy from the
interior of the heated cavities into the surrounding cold gas
(\eg Soker 2004; Sternberg \etal 2007).   This
means that while in the pure-hydro run, the bubbles deposit most of
their internal energy at the resolution-dependent radius at which
they get disrupted, as indicated  by the spike in temperature at $R
\approx 30$ kpc. Hence, the turbulent motions captured by the subgrid model
cause much more gradual heating, which leads to shallow temperature
and entropy profiles at all radii $\leq 30$ kpc.

At the same time, the well-mixed clouds in the subgrid-turbulence run
remain coherent to larger radii, increasing the sphere of influence
over which AGN-driven heating is effective.  This can be clearly seen
in a comparison of the metal profiles between the two runs, which
indicates enhancement of almost an order of magnitude  in the
subgrid-turbulence runs relative to the pure-hydro runs, an
enhancement which is particularly strong at large radii, but
significant even in the central regions of the cluster.

\begin{figure*}[ht]
\centerline{\includegraphics[height=13.0cm]{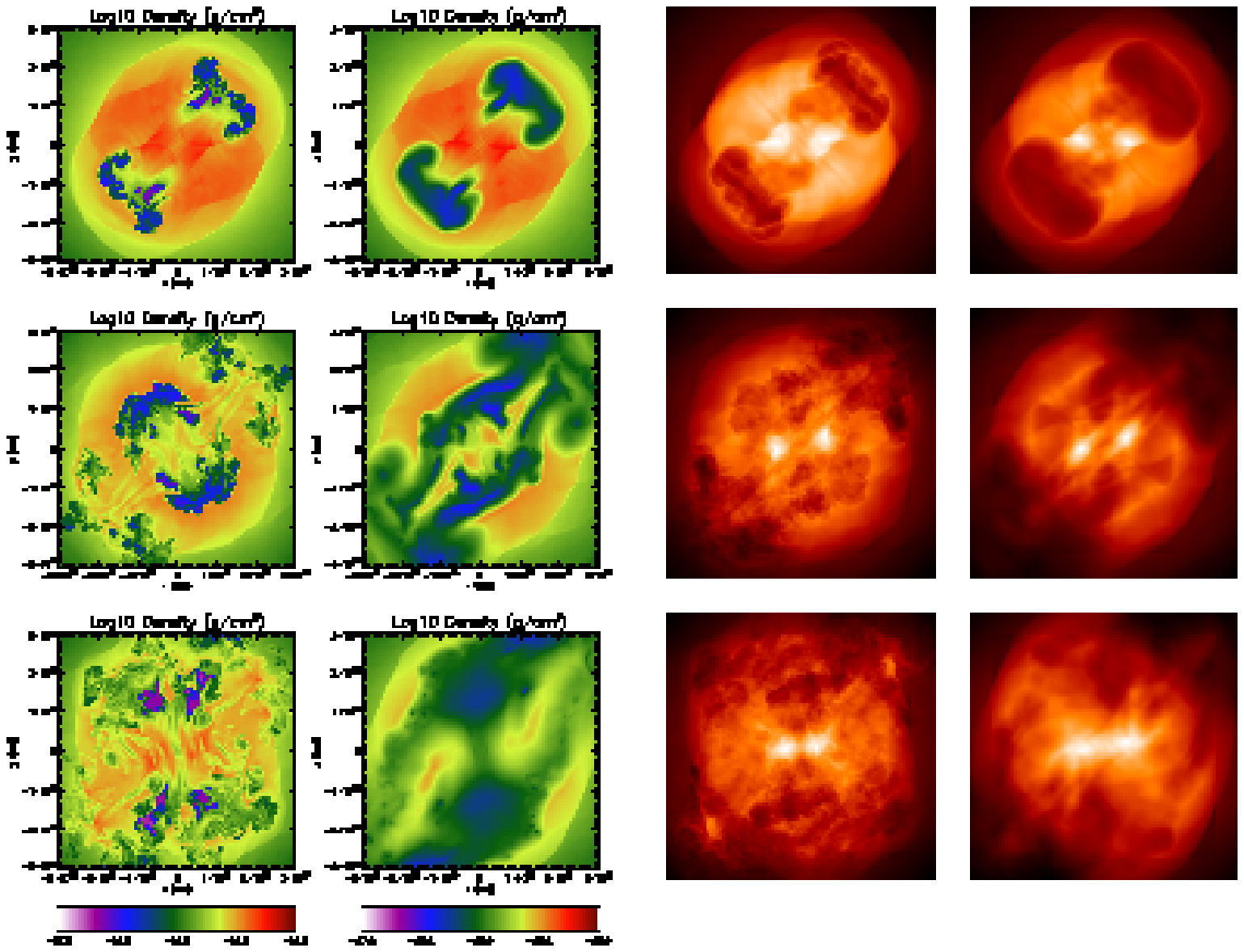}}
\caption{Snapshots of mean-flow quantities in the periodic Sedov
bubble runs at 100 Myrs, 200 Myrs, and 300 Myrs (arranged from top to
bottom in each column).   Panels and symbols are as in Figure
\protect\ref{fig:periodicmean}.}
\label{fig:periodicmeanshock}
\end{figure*}

\begin{figure}[ht]
\centerline{\includegraphics[height=9.0cm]{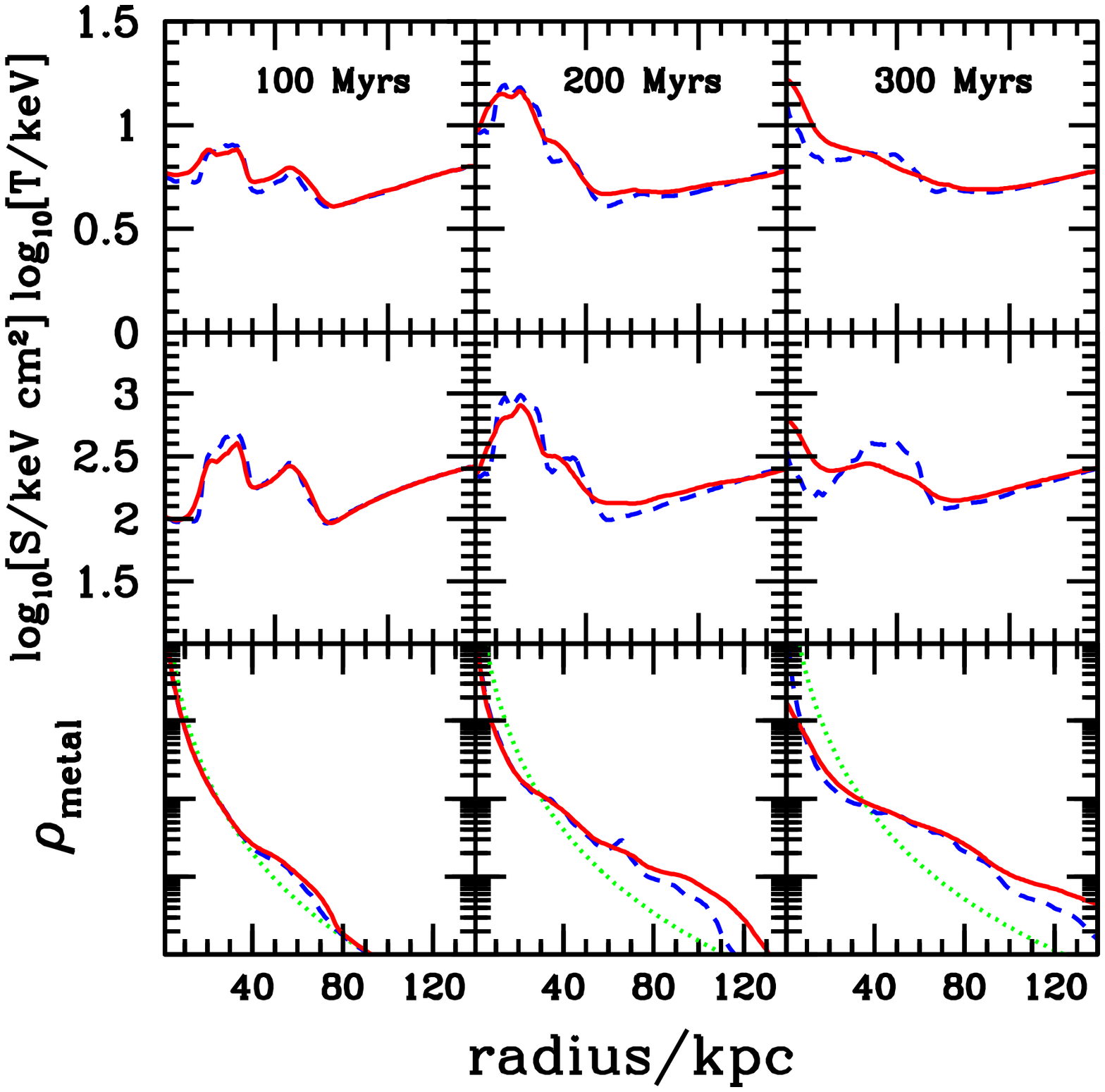}}
\caption{Radial profile of mean-flow quantities in Sedov
multiple-bubble  runs at times of 100 Myrs, 200 Myrs, and 300 Myrs,
arranged in columns from left to right.  From top to bottom the rows
show the average temperature, entropy, and metal density.  Symbols are
as in Figure 16.\\}
\label{fig:radialSR}
\end{figure}

Turning to the Sedov-bubble case, similar trends are apparent.  Here,
an extremely large number of AMR zones  were necessary at  late times
($\approx 10^5$ blocks of $8^3$ zones each), forcing us to end our
simulations at 300 Myrs.  However, the density slices and X-ray images
presented in  Figure \ref{fig:periodicmeanshock} over this time period
paint a similar picture to the evacuated-bubble runs.  Again the
build-up of turbulent motions over time causes the properties of the
two simulations to diverge.  Again this increase in mixing is
especially notable near the core, where a large number of small
fragments in the pure-hydro case are replaced by an  overall smooth
and quadrapole-like configuration in the subgrid-turbulence run.  And
again the subgrid model serves to damp out much of the chaotic
amplification of small assymetries.  Furthermore,
the X-ray images of the subgrid turbulence runs show larger and
smoother cavities, although interestingly, the pure-hydro runs contain
significant associations of shredded bubble material that appear as
coherent X-ray depressions, albeit somewhat more ragged depressions than are
seen in the observations.

As in the evacuated-bubble case, the radial profiles plotted in Figure
\ref{fig:radialSR} show that subgrid turbulence leads to more
efficient heating in the central $R \leq 30$ kpc core.  On the other
hand, at larger radii heating is dominated by shocks and is similar
between the two runs.  Note that as $E_{\rm expand}$ is substantially
greater than $ E_{\rm buoyancy} \approx E_{\rm cool},$ these
simulations show an overall increase in the central temperature and
entropy  of the cluster at late times.  In reality, entropy is
unlikely to rise beyond the point at which the cooling near the core
becomes inefficient, as this would halt gas accretion onto the central
black hole and the further creation of bubbles.  Accounting for this
change in gas accretion rate would lead to a self-regulating
configuration (\eg Voit \& Bryan 2001; 
 B\^irzan \etal 2004;  McNamara \& Nulsen 2007; Voit
\etal 2008) whose modeling falls beyond the scope of our study here.

In the lower panels of this figure, we plot the metal profiles between
the two runs. Although metal mixing is somewhat more efficient at
larger radial in the  subgrid-turbulence run,  the metal profiles are
much more similar between the two runs than they were in the the
evacuated-bubble case.
 
\begin{figure*}[ht]
\centerline{\includegraphics[height=10.0cm]{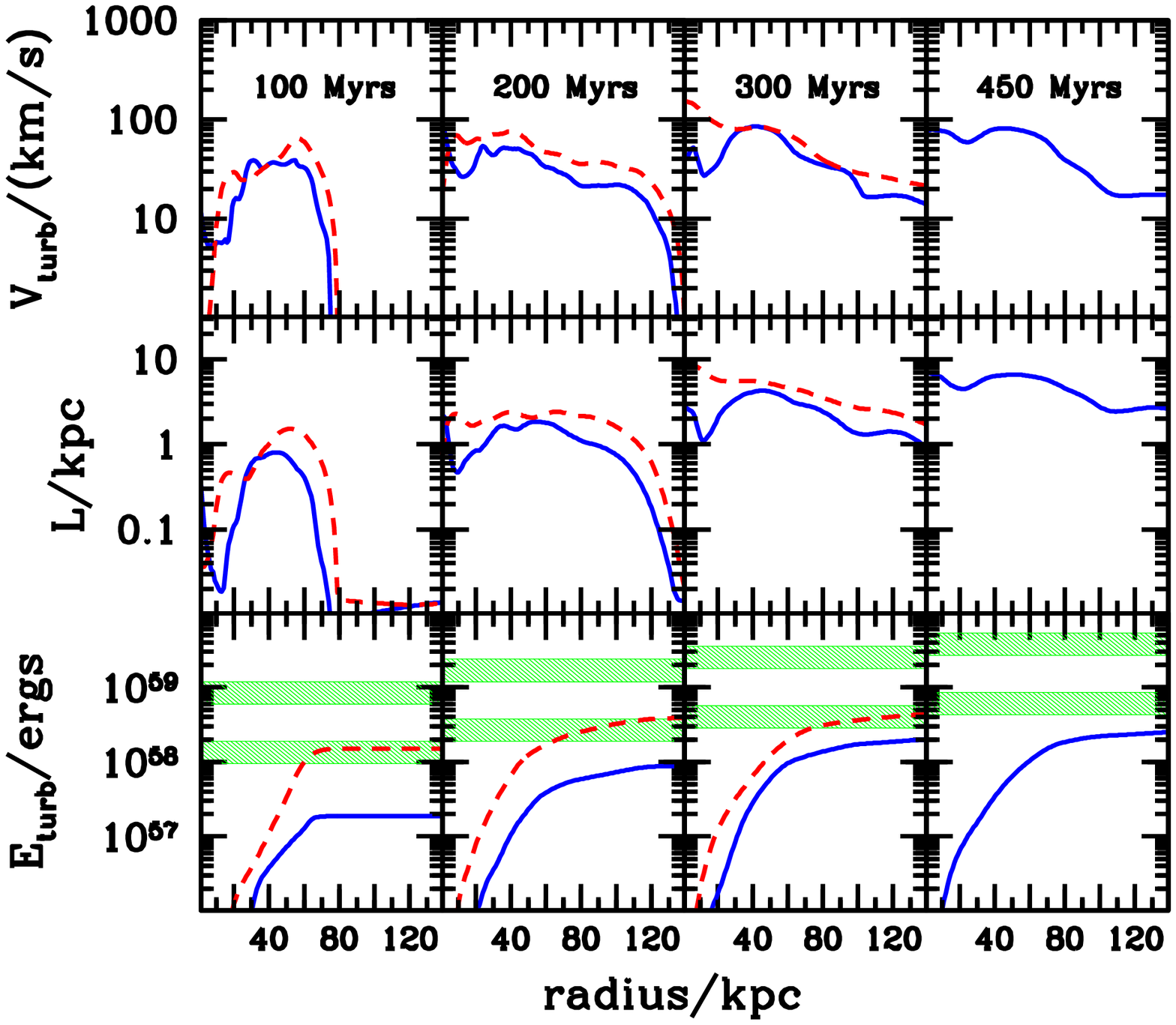}}
\caption{Radial profile of turbulence quantities in  the evacuated
multiple bubble run (5DCER, solid lines), and the Sedov
multiple-bubble run (5DCSR, dashed lines), at times of 100 Myrs, 200
Myrs, 300 Myrs, and 450  Myrs, arranged in columns from left to right.
From top to bottom the rows show the average turbulent velocity and
txurbulent  length scale as a function of
radius, and the total turbulent kinetic energy within this radius.  As
in Figure  \protect\ref{fig:radialSS} the shaded regions in the lower
panels show $1-2$ of  $E_{\rm buoyancy}$ and $E_{\rm expand} + E_{\rm
buoyancy}.$\\}
\label{fig:radialTR}
\end{figure*}

Finally, in Figure \ref{fig:radialTR} we compare the subgrid
turbulence  quantities between the evacuated repeated bubble run
(5DCER) and the Sedov repeated bubble run (5NCER).   As in the single
bubble runs turbulent velocities are only slightly increased by the RM
instability, even though the total energy input in this case is
substantially higher.  Thus $V_{\rm turb} \approx 100-200$ km/s in
both  runs, roughly $15-30\%$ of the sound speed of the cluster.
Although this  is somewhat smaller than the Mach number of $\approx
0.5$ suggested by the resonance line studies of Perseus by Churazov
\etal (2005), again our  model only provides a lower limit on this
value as it does not include all mechanisms that drive turbulence.

Comparing the turbulent length scale between the two runs, we again
find that the enhancement in the Sedov bubble case is not dramatic.
Apart from a noticeable increase in $L$ at $R \lesssim 20$ kpc in the
Sedov bubble runs, both runs achieve similar values of $L$, which
averages $\approx 5$ kpc.  Combining $L$ and $V$ to construct a
diffusion coefficient gives typical values of $500$ kpc km/s, which
are in excellent agreement with the effective diffusion coefficient
inferred by the abundance profiles studies of Perseus by Rebusco \etal
(2005), again  suggesting that more complete self-regulating models
with subgrid turbulence could do well at explaining this and other
clusters.

In all cases, the total turbulent energy, shown in the the bottom
panels, is $\approx$ 1-2\% of the buoyant energy in the bubbles, and
even a smaller fraction of $E_{\rm buoyancy}+E_{\rm expand}.$ Thus, as
suggested above, turbulent dissipation is not likely to be an
important factor.  Rather it is the turbulent mixing of heated and
enriched gas that plays a fundamental role in the evolution of
AGN-heated galaxy clusters.

\section{Conclusions}

A wide range of observations suggest that AGN-feedback  plays a key
role in the evolution of cool-core galaxy clusters.  At the same time,
theoretical studies have pointed out some of the many physical
mechanisms that may be important in this evolution, including
viscosity, magnetohydrodynamic effects, heat conduction, and
cosmic-ray heating.  While any or all of these may operate in nature,
none of them can be fully understood without first accurately
capturing the underlying hydrodynamic evolution of the ICM.  It is
with this basic goal in mind that we have used a subgrid turbulence model
to study AGN heating in  an inviscid fluid, neglecting other effects
such as magnetic fields and heat-conduction.

Within this context, our study has been focused on two key issues: the
growth of instabilities and turbulence caused by AGN-heated bubbles
and the role of these instabilities in determining the evolution of
the bubbles and the surrounding medium.  Clearly, there  are other
sources of turbulence that might add  similar  levels of turbulent
energy into the ICM, such as mergers of large subclusters or motions
of galaxies within a cluster.  Likewise, the DT06 subgrid turbulence
model that we have employed does not include all processes that
generate turbulence, such as the shear-driven Kelvin-Helmholtz
instability.  However, our tests show that it is effective in
capturing the growth of the extremely important RT and RM
instabilities.  Thus
although many aspects of cluster evolution remain uncertain and beyond
the scope of this work, there are a number of robust conclusions that
we can make about the role of these two instabilities in shaping the
evolution and impact of AGN-heated cavities in clusters.  In
particular we find that:
 
\begin{itemize}

\item  Many of the RT and RM unstable modes that drive the evolution
of the bubbles evolve on scales that are far below the resolution
limits of current simulations. The superposition of these unstable
modes  smears out the interface between the bubbles and the ambient
medium, transforming them into clouds of mixed material that stay
intact and expand as they rise in the  stratified cluster medium.
This mixing can explain the coherent X-ray cavities detected in
clusters of galaxies.
The subgrid-turbulence model also greatly reduces the sensitivity of our results on the resolution of the computational grid and the detailed choice of initial conditions.

\item  Within the clouds, turbulent motions quickly attain typical
velocities of $\approx 200$ km s$^{-1},$ roughly $10\%$ of the
internal sound speed  and $20\%$ of the sound speed of the surrounding
ICM.  Similarly, the scale of the turbulent eddies rises swiftly but
does not exceed the $\approx 30$ kpc scale of clouds.  A
typical turbulent diffusivity is then $\approx 500$ km/s kpc, which is in excellent agreement with the diffusion coefficient inferred by the abundance profiles studies of Perseus by
Rebusco \etal (2005).

\item Subgrid turbulence is likely to enhance metal transport
significantly.  In our fiducial single-bubble evacuated pure-hydro
run, metal transport is halted at $\approx 50$ kpc by bubble
disruption, which occurs when RT instabilities have shredded the
evacuated region into resolution-limited cavities.  Yet, in the
subgrid-turbulence run, small-scale fluctuations act to keep the cloud
coherent for much longer, thus causing it to distribute metals
out to larger radii.

\item In the case where the bubbles produce shock waves, the turbulent
velocities and length scales are systematically enhanced with
respect to the evacuated-bubble run, due to the RM
instability.  While such $\approx 300$ km/s velocities are roughly
50\% greater than in the runs with evacuated bubbles, they are still
well below the $\approx 500$ km/s radial velocities of the clouds.
However they are large enough to be probed by studying the emission lines 
of heavy ions with the future {\em Constellation-X} satellite, which will have
an envisaged spectral resolution of 1-2 eV.

\item Calculating the turbulent kinetic energy produced by the rising
bubbles, we find that this is only $\approx 1\%$ of  the total energy
available for the bubbles to  heat the cluster.  Hence, we do not
expect the energy of the turbulent motions themselves to play a major
role in the heating of the cool-core regions of the ICM.  Rather, the
main role of turbulence is to increase the efficiency with which the
thermal energy of the rising clouds is mixed into the surrounding  gas.

\item   Finally, runs that include radiative cooling and multiple
episodes of AGN-feedback indicate that the  impact of turbulence
continues to increase with successive generations of heating.   This
is because turbulence driven by previous feedback events remains
behind, enhancing the mixing of subsequent bubbles.  While in our
pure-hydro runs, the bubbles deposit most of their internal energy  at
the resolution-dependent radius at which they are disrupted, the
turbulent motions captured by the subgrid model lead to more gradual
heating, correspondingly shallower temperature and entropy profiles,
and shallower metal gradients.

\end{itemize}

In summary, the properties, evolution, and appearance of AGN-blown
bubbles in clusters are substantially different when one properly
accounts for turbulent motions on scales well below the limits of
current pure-hydro simulations.  Although accounting for these motions
may not provide the ultimate solution to many of the mysteries
surrounding galaxy clusters, it is a crucial step forward in the
modeling of feedback and the interaction between AGN and the
ICM. Subgrid models such as the one developed in DT06 provide us with
tools for capturing this physics.  In fact, the numerical methodology
presented here is likely to have applications in other areas of
astrophysics where hydrodynamic modeling of RT instabilities is
crucial, such as supernovae, supernova remnants, and galactic winds.
It is clear that while many physical processes may play important
roles in clusters and other environments, these can only be understood
when carefully disentangled from the impact of subgrid turbulence.

\acknowledgments

We are grateful to Baolian Cheng, Rolf Jansen, A.J. Scannapieco, Rob
Thacker,  Frank Timmes,  and Dean Townsley for their  many useful
comments and suggestions and to the referee, Noam Soker, for his careful
reading of our manuscript.  MB acknowledges the support by the DFG
grant BR 2026/3 within the Priority Programme ``Witnesses of Cosmic
History'' and the supercomputing grants NIC 2195 and 2256 at the
John-Neumann Institut at the Forschungszentrum J\"ulich.  All
simulations were conducted on the ``Saguaro'' cluster operated by the
Fulton School of Engineering at Arizona State University.  The results
presented here were produced using the FLASH code, a product of the
DOE ASC/Alliances-funded Center for Astrophysical Thermonuclear
Flashes at the University of Chicago.


\end{document}